\documentclass[twocolumn]{aastex631}

\usepackage{graphicx}

\shorttitle{AGE-PRO X. Dust Substructures, Disk Geometries, and Dust-disk Radii}
\shortauthors{M. Vioque, N. T. Kurtovic, L. Trapman, and et al.}
\let\bfseries\mdseries
\graphicspath{{./}{figures/}{figures/figures_archive/}}

\begin{document}

\title{The ALMA Survey of Gas Evolution of PROtoplanetary Disks (AGE-PRO): X. Dust Substructures, Disk Geometries, and Dust-disk Radii}

\author[0000-0002-4147-3846]{Miguel Vioque}
\affiliation{European Southern Observatory, Karl-Schwarzschild-Str. 2, 85748 Garching bei München, Germany} 
\affiliation{Joint ALMA Observatory, Alonso de C\'ordova 3107, Vitacura, Santiago 763-0355, Chile} 
\author[0000-0002-2358-4796]{Nicolás T. Kurtovic}
\affiliation{Max-Planck-Institut f\"ur Extraterrestrische Physik, Giessenbachstrasse 1, 85748 Garching, Germany}
\author[0000-0002-8623-9703]{Leon Trapman}
\affiliation{Department of Astronomy, University of Wisconsin-Madison, 475 N Charter St, Madison, WI 53706, USA}
\author[0000-0002-5991-8073]{Anibal Sierra}
\affiliation{Departamento de Astronom\'ia, Universidad de Chile, Camino El Observatorio 1515, Las Condes, Santiago, Chile}
\affiliation{Mullard Space Science Laboratory, University College London, Holmbury St Mary, Dorking, Surrey RH5 6NT, UK}
\author[0000-0002-1199-9564]{Laura M. Pérez}
\affiliation{Departamento de Astronom\'ia, Universidad de Chile, Camino El Observatorio 1515, Las Condes, Santiago, Chile}
\author[0000-0002-0661-7517]{Ke Zhang}
\affiliation{Department of Astronomy, University of Wisconsin-Madison, 475 N Charter St, Madison, WI 53706, USA}
\author[0000-0003-2045-2154]{Pietro Curone}
\affiliation{Departamento de Astronom\'ia, Universidad de Chile, Camino El Observatorio 1515, Las Condes, Santiago, Chile}
\affiliation{Dipartimento di Fisica, Università degli Studi di Milano, Via Celoria 16, I-20133 Milano, Italy}
\author[0000-0003-4853-5736]{Giovanni P. Rosotti}
\affiliation{Dipartimento di Fisica, Università degli Studi di Milano, Via Celoria 16, I-20133 Milano, Italy}
\author[0000-0003-2251-0602]{John Carpenter}
\affiliation{Joint ALMA Observatory, Alonso de C\'ordova 3107, Vitacura, Santiago 763-0355, Chile}
\author[0000-0002-1103-3225]{Benoît Tabone}
\affiliation{Université Paris-Saclay, CNRS, Institut d'Astrophysique Spatiale, Orsay, France}
\author[0000-0001-8764-1780]{Paola Pinilla}
\affiliation{Mullard Space Science Laboratory, University College London, Holmbury St Mary, Dorking, Surrey RH5 6NT, UK}
\author[0000-0003-0777-7392]{Dingshan Deng}
\affiliation{Lunar and Planetary Laboratory, the University of Arizona, Tucson, AZ 85721, USA}
\author[0000-0001-7962-1683]{Ilaria Pascucci}
\affiliation{Lunar and Planetary Laboratory, the University of Arizona, Tucson, AZ 85721, USA}

\author[0000-0002-1575-680X]{James Miley}
\affiliation{Departamento de Física, Universidad de Santiago de Chile, Av. Victor Jara 3659, Santiago, Chile}
\affiliation{Millennium Nucleus on Young Exoplanets and their Moons (YEMS), Chile}
\affiliation{Center for Interdisciplinary Research in Astrophysics and Space Exploration (CIRAS), Universidad de Santiago, Chile}

\author[0000-0002-7238-2306]{Carolina Agurto-Gangas}
\affiliation{Departamento de Astronom\'ia, Universidad de Chile, Camino El Observatorio 1515, Las Condes, Santiago, Chile}

\author[0000-0002-2828-1153]{Lucas A. Cieza}
\affiliation{Millennium Nucleus on Young Exoplanets and their Moons (YEMS), Chile}
\affiliation{Instituto de Estudios Astrofísicos, Universidad Diego Portales, Av. Ejercito 441, Santiago, Chile}

\author[0009-0004-8091-5055]{Rossella Anania}
\affiliation{Dipartimento di Fisica, Università degli Studi di Milano, Via Celoria 16, I-20133 Milano, Italy}

\author[0000-0003-3573-8163]{Dary A. Ruiz-Rodriguez}
\affiliation{Joint ALMA Observatory, Alonso de C\'ordova 3107, Vitacura, Santiago 763-0355, Chile}

\author[0000-0003-4907-189X]{Camilo Gonz\'alez-Ruilova}
\affiliation{Instituto de Estudios Astrofísicos, Universidad Diego Portales, Av. Ejercito 441, Santiago, Chile}
\affiliation{Millennium Nucleus on Young Exoplanets and their Moons (YEMS), Chile}
\affiliation{Center for Interdisciplinary Research in Astrophysics and Space Exploration (CIRAS), Universidad de Santiago, Chile}

\author[0000-0001-9961-8203]{Estephani E. TorresVillanueva}
\affiliation{Department of Astronomy, University of Wisconsin-Madison, 475 N Charter St, Madison, WI 53706, USA}

\author[0000-0002-6946-6787]{Aleksandra Kuznetsova}
\affiliation{Center for Computational Astrophysics, Flatiron Institute, 162 Fifth Ave., New York, New York, 10025, USA}
\affiliation{Department of Physics, University of Connecticut, 196A Auditorium Road, Unit 3046, Storrs, CT 06269, USA}

\begin{abstract}

We perform visibility fitting to the dust continuum Band 6 1.3 mm data of the 30 protoplanetary disks in the AGE-PRO ALMA Large Program. We obtain disk geometries, dust-disk radii, and azimuthally symmetric radial profiles of the intensity of the dust continuum emission. We examine the presence of continuum substructures in the AGE-PRO sample \textbf{by using these radial profiles} and their residuals. We detect substructures in 15 out of 30 disks. We report five disks with large ($>$15 au) inner dust cavities. The Ophiuchus Class I disks show dust-disk substructures in $\sim$80\% of the resolved sources. This evidences the early formation of substructures in protoplanetary disks. A spiral is identified in IRS 63, hinting to gravitational instability in this massive disk. We compare our dust-disk brightness radial profiles with gas-disk brightness radial profiles and discuss colocal substructures in both tracers. In addition, we discuss the evolution of dust-disk radii and substructures across Ophiuchus, Lupus, and Upper Scorpius. We find that disks in Lupus and Upper Scorpius with large inner dust cavities have typical gas-disk masses, suggesting an abundance of dust cavities in these regions. The prevalence of pressure dust traps at later ages is supported by a potential trend with time with more disks with large inner dust cavities (or \textit{transition disks}) in Upper Scorpius \textbf{and the absence of evolution of dust-disk sizes with time in the AGE-PRO sample.} We propose this is caused by an evolutionary sequence with a high fraction\textbf{ of protoplanetary disks} with inner protoplanets \textbf{carving dust cavities}.







\end{abstract}
\section{Introduction}\label{sec:intro}

The millimeter continuum emission of protoplanetary disks allow us to study the millimeter sized dust component of the disk midplanes, and its evolution over time (see \citealp{2020ARA&A..58..483A,2023ASPC..534..423B, Manara_2023_PPVII,2023ASPC..534..501M}, and references therein). \citet{2017ApJ...845...44T} and \citet{2018ApJ...865..157A} found that the millimeter continuum luminosity scales with the disk area, indicating a large contribution to the luminosity from optically thick emission. Nevertheless, millimeter continuum luminosities have often been used to estimate a total dust-disk mass, provided some assumptions on average disk temperature, dust opacity, and optically thin emission (e.g. \citealp{2013ApJ...771..129A, 2016ApJ...831..125P, 2020A&A...640A..19T}). These masses have in turn been used extensively to probe the evolution of the mass reservoir in protoplanetary disks as they disperse, and material is accreted, ejected, or locked into planets (e.g. \citealp{2016A&A...591L...3M,2018ApJS..238...19T,2022A&A...661A..53V,2023A&A...679A..82M}). If multiple millimeter wavelengths are considered, continuum thermal emission can also be used to constrain grain growth and drift (e.g. \citealp{2014prpl.conf..339T, 2021ApJS..257...14S}). 


From the  millimeter continuum emission of protoplanetary disks, dust-disk radii can also be measured. However, it has proved challenging to correctly define an outer radius in protoplanetary disks, and often radii enclosing a certain percentage of the total flux are used (see \citealp{2023ASPC..534..501M}, and references therein). Dust-disk radii have also been used extensively to describe disk populations (e.g. \citealp{2019MNRAS.486.4829R, 2020ApJ...895..126H, 2021A&A...649A..19S}) and to characterize internal and environmental processes of protoplanetary disk evolution.


Another focus of dust continuum emission studies is on dust substructures, and their relation to planet formation mechanisms (see \citealp{2023ASPC..534..423B}, and references therein). Dust-disk substructures are diverse and seem ubiquitous whenever disks are observed with enough resolution (\citealp{2018ApJ...869...17L, 2023ASPC..534..423B,2023ApJ...952..108Z}). Hence, these studies have been particularly enhanced by Atacama Large Millimeter/submillimeter Array (ALMA) high-resolution observations (\citealp{Andrews_DSHARP_2018ApJ...869L..41A, 2020ARA&A..58..483A}). Disk substructures have been studied to great detail in several objects (e.g. \citealp{2015ApJ...808L...3A, 2019A&A...625A.118K, 2021A&A...648A..33M, 2024A&A...686A.298G}), but have also been used as a tracer of evolution in populations of protoplanetary disks (e.g. \citealp{2018ApJ...860...77E, 2019ApJ...882...49L, 2023ApJ...951....8O, 2024PASJ...76..437Y}). Axisymmetric structures (rings, gaps) appear to be more common than asymmetric ones (spirals, crescents). It is known that dust substructures play a major role in planet formation and disk evolution mechanisms, but their variety and complexity indicate they are sensitive to several phenomena and have different origins, and are often challenging to reconcile with theoretical predictions (\citealp{2023ASPC..534..423B}).


As part of the ALMA Large Program AGE-PRO (the ALMA Survey of Gas Evolution of PROtoplanetary Disks, \citealp{AGEPRO_I_overview}, 2021.1.00128.L) we dedicate this paper to the analysis of the Band 6 1.3 mm dust continuum emission of the AGE-PRO sample of 30 protoplanetary disks, studying evolution across the three considered regions: 10 disks in Ophiuchus, 10 disks in Lupus, and 10 disks in Upper Scorpius, all in a narrow stellar mass range (0.3-0.8 $M_{\odot}$). 

We follow the same methodology homogeneously for all disks, which can be summarized in three steps. First (Sect. \ref{sec:methodology}), we obtain disk inclinations and position angles (PA). In a second step (Sect. \ref{Frank}), we fit the visibilities of the dust continuum emission using as the input the geometries found in the first step. The resulting models provide us with azimuthally symmetric radial intensity profiles of the continuum emission. We use these radial profiles to identify dust substructures (Sect. \ref{Frank_subsec}) and to obtain curves of growth from which to extract dust-disk radii (Sect. \ref{sec_disk_radii}). In the last step (Sect. \ref{residuals}), we analyze the residuals of the best model for each source to identify nonazimuthally symmetric structure. \textbf{AGE-PRO observations have an angular image resolution of $\sim0.15-0.30$", but via visibility fitting, we achieve smaller resolution elements (Table \ref{Table2}). The average continuum sensitivity of the survey is 0.025 mJy/beam rms. More information on the data can be found in the AGE-PRO papers of the three regions: \citet{AGEPRO_II_Ophiuchus}, \citet{AGEPRO_III_Lupus}, and \citet{AGEPRO_IV_UpperSco}.}

We analyze and discuss our results in Sect. \ref{analysis}. We start by placing the AGE-PRO sample of disks into context with respect to the broader population of known protoplanetary disks. We then describe trends of the dust-disk radii with age and between star forming regions (Sect. \ref{Sec_population}). We analyze the evolution and type of dust substructures at different ages and in different regions in Sects. \ref{discuss_structures} and \ref{Oph_substruc}. We conclude the analysis with a comparison of the radial intensity profiles of the dust continuum emission reported in this work with the gas-disk radial profiles derived in \citet[Sect. \ref{dust-disk vs. gas-disk}]{AGEPRO_XI_gas_disk_sizes}. We summarize our conclusions in Sect. \ref{Conclusions2}.













\begin{figure*}[ht!]
\includegraphics[width=\textwidth]{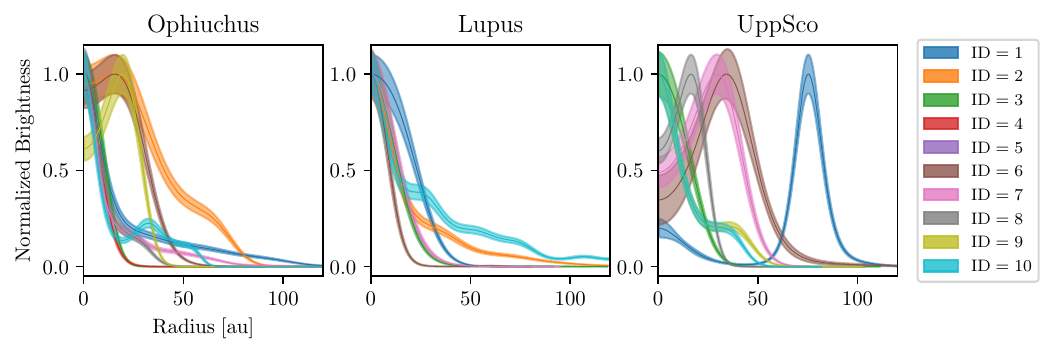}
\caption{Peak normalized brightness radial profiles obtained from the best \texttt{Frankenstein} fit of each target and only for the well-resolved disks in each region. Uncertainties are derived from the flux calibration error, \texttt{Frankenstein} error, and deviations from azimuthal symmetry. Colors indicate source number (see Table \ref{Table1} for naming convention).}\label{Plot: Three_regions}
\end{figure*} 

\section{Geometries of AGE-PRO Disks}\label{sec:methodology}

%



We dedicate this section to characterize the disk geometries (i.e. inclination and PA) of the AGE-PRO sample from the continuum visibilities (Sect. \ref{galario}). For those disks with marginally resolved continuum visibilities, we try to obtain disk geometries from the \textsuperscript{12}CO moment zero maps (Sect. \ref{Leon_work}).



\subsection{Gaussian Visibility Fitting}\label{galario}

In this section, we obtain disk geometries and phase offsets by fitting the interferometric continuum visibilities using \texttt{Galario} (\citealp{2018MNRAS.476.4527T}), a Python library that compares observed visibilities with model visibilities in the visibility plane (abbreviated as uv-plane). The Band 6 (\citealp{2014ITTST...4..201K}) visibilities are extracted from the self-calibrated, phase-centered, and line-free AGE-PRO measurement sets using a modified version of the \textit{export uvtable} function in \citet{2017zndo...1003113T}. In our version, the uv-distances are normalized using the wavelength of each channel and spectral window, instead of the average wavelength of the whole data set. 

Using \texttt{Galario}, we fit with Markov Chain Monte Carlo (MCMC; \texttt{emcee}; \citealp{2013PASP..125..306F}) the observed {continuum} visibilities of each disk with an azimuthally symmetric Gaussian function:

\begin{equation}\label{Eq.1}
I = f_{0}\,e^{-0.5(R/\sigma)^{2}}.  
\end{equation}

Disk inclinations (in the range $0^{\circ}$ to $90^{\circ}$), disk PA (in the range $0^{\circ}$ to $180^{\circ}$), and offsets with respect to the phase center ($\Delta$RA and $\Delta$Dec) are part of the MCMC posterior. For the MCMC, we used ensembles of 40 walkers and 6000 steps (of which 2000 are burned-in). Some sources required small deviations from these ensembles for better convergence. We note that for 2MASS J16202863-2442087 (Upper Scorpius 7) we only use observations of quiescent continuum emission (see \citealp{AGEPRO_XII_mm_var_USco7}). 





A number of AGE-PRO sources have companions within the field of view, which were masked during the data calibration (\citealp{AGEPRO_IV_UpperSco,AGEPRO_III_Lupus,AGEPRO_II_Ophiuchus}). A fraction of those are likely binary sources (e.g. Sz 65 and Sz 66, \citealp{2024A&A...682A..55M}). However, these companions have negligible effects on the visibility modeling due to their considerable faintness and their substantial distance from the main source.


{There are eight disks that are marginally resolved, \textbf{leading to higher uncertainties} in the Gaussian fit of their dust visibilities. These sources are Ophiuchus 5, Lupus 4, 5, 8, and 9, and Upper Scorpius 2, 4, and 5 (see Table \ref{Table1} for naming convention). We consider these sources as poorly resolved in continuum visibility space. A different approach to try to constrain the geometries of these sources from the gas is presented in Sect. \ref{Leon_work}.


The results for the geometry and phase offset of each disk from the \texttt{Galario} {plus MCMC fit} are summarized in Table \ref{Table1}. Uncertainties of each parameter indicate the 16\% and 84\% percentiles of the posterior distribution. {For AGE-PRO internal consistency}, in the case of 2MASS J16120668-3010270 (Upper Scorpius 1), we adopt the geometries identified in \citet{2024ApJ...974..102S} using a similar procedure to the one of this work.

\subsection{Sources with Geometries from the Gas-disk}\label{Leon_work}

For the eight sources that are marginally resolved in their continuum visibilities, we consider a different approach to constrain their disk geometries. Gas emission is often more extended than the dust, thereby providing an additional and often more reliable constraint on the disk geometry in these cases where the dust-disks are only marginally resolved. Using the method described in \citet{AGEPRO_XI_gas_disk_sizes}, \textbf{we fit two-dimensional Gaussians to the \textsuperscript{12}CO moment zero maps of these sources (except for Ophiuchus 5, which shows too much \textsuperscript{12}CO cloud contribution for this methodology to work).} The details of the results obtained with these fits to the \textsuperscript{12}CO moment zero maps are presented in Appendix \ref{Leon_work_appendix}.

In summary, of the seven Lupus and Upper Scorpius marginally resolved sources in the continuum visibilities (Sect. \ref{galario}), we consider in this work the \textsuperscript{12}CO-based geometries for Lupus 4, 5, and Upper Scorpius 5. The other sources show nonconclusive fits to the \textsuperscript{12}CO moment zero maps.  Even though it is well resolved in continuum emission, for Upper Scorpius 3, we also use the \textsuperscript{12}CO-based geometries as they provide much better residuals (see Sect. \ref{residuals}). Table \ref{tab: gasfit disk geometry} in Appendix \ref{Leon_work_appendix} details the sources for which \textsuperscript{12}CO-based geometries were adopted, and shows the comparison with the dust-based geometries obtained for them in Sect. \ref{galario}.

The final set of geometries (PA and inclinations) and phase offsets for the AGE-PRO sample of protoplanetary disks is presented in Table \ref{Table1}. All phase offsets come from the \texttt{Galario} fit of Sect. \ref{galario}. The geometries derived in Sect. \ref{galario} and \ref{Leon_work} are adopted for the visibility fitting to follow (Sect. \ref{Frank}).

\begin{deluxetable*}{llcccccc}
\tablecaption{Geometries and Phase Offsets of AGE-PRO Disks \label{Table1}}
\tablewidth{0.45\textwidth}
\tablehead{
\colhead{AGE-PRO ID} &
\colhead{Source} & 
\colhead{RA} & \colhead{Dec} & \colhead{PA} &
\colhead{Inc} &
\colhead{$\Delta$RA} & \colhead{$\Delta$Dec} \\
\colhead{} & \colhead{} & \colhead{(hh:mm:ss.ss)} & 
\colhead{(dd:mm:ss.s)} & 
\colhead{(deg)} & \colhead{(deg)} & \colhead{(mas)} &
\colhead{(mas)} 
}
\startdata
Ophiuchus  \\
\hline
  1 (ISO-Oph 37) & SSTc2d J162623.6-242439 & 16h26m23.57s & -24d24m40.1s &      $48.43^{+0.02}_{-0.02}$ &     $72.59^{+0.02}_{-0.02}$ &     $-0.48^{+0.11}_{-0.11}$ &    $-2.51^{+0.09}_{-0.10}$  \\
  2 (ISO-Oph 94) & SSTc2d J162703.6-242005 & 16h27m03.59s & -24d20m05.5s &      $50.18^{+0.07}_{-0.06}$ &     $76.40^{+0.07}_{-0.07}$ &   $1.65^{+0.31}_{-0.31}$ &    $0.19^{+0.26}_{-0.26}$ \\
  3 (ISO-Oph 129) & SSTc2d J162719.2-242844 & 16h27m19.20s & -24d28m44.5s &      $91.58^{+0.27}_{-0.28}$ &   $61.25^{+0.25}_{-0.25}$ &     $5.04^{+0.15}_{-0.14}$ &    $-4.76^{+0.10}_{-0.11}$ \\
  4 (Elias 2-32) & SSTc2d J162728.4-242721 & 16h27m28.44s & -24d27m21.8s &      $100.7^{+2.5}_{-2.6}$ &     $61.0^{+2.2}_{-2.2}$ &     $3.4^{+1.3}_{-1.3}$ &    $-7.50^{+0.87}_{-0.85}$\\
  5 (ISO-Oph 161) & SSTc2d J162737.2-244237 & 16h27m37.24s & -24d42m38.5s &      $58^{+86}_{-34}$ &   $25^{+16}_{-13}$ &     $4.04^{+0.36}_{-0.38}$ &    $-1.92^{+0.31}_{-0.32}$ \\
  6 (ISO-Oph 165) & SSTc2d J162738.9-244020 & 16h27m38.93s & -24d40m21.2s &      $167.85^{+0.08}_{-0.08}$ &   $74.34^{+0.12}_{-0.11}$ &     $6.53^{+0.14}_{-0.14}$ &    $-2.29^{+0.19}_{-0.19}$ \\
  7 (IRS 63)  & SSTc2d J163135.6-240129 & 16h31m35.65s & -24d01m30.1s & $149.56^{+0.04}_{-0.03}$ & $46.94^{+0.02}_{-0.03}$ & $-0.79^{+0.04}_{-0.04}$ & $-0.90^{+0.04}_{-0.04}$ \\
  8 & 2MASS J16230544-2302566 & 16h23m05.42s & -23d02m57.6s &      $72.50^{+0.54}_{-0.56}$ &     $71.84^{+0.55}_{-0.57}$ &     $-10.03^{+0.98}_{-0.99}$ &    $-8.69^{+0.63}_{-0.62}$\\
  9 (ISO-Oph 43) & SSTc2d J162627.5-244153 & 16h26m27.54s & -24d41m54.0s &      $159.55^{+0.30}_{-0.29}$ &   $73.48^{+0.45}_{-0.43}$ &     $-4.96^{+0.45}_{-0.48}$ &    $-1.14^{+0.61}_{-0.61}$ \\
 10 (ISO-Oph 127) & SSTc2d J162718.4-243915 & 16h27m18.37s & -24d39m15.3s &      $52.918^{+0.090}_{-0.090}$ &   $73.617^{+0.093}_{-0.089}$ &     $0.89^{+0.31}_{-0.32}$ &    $-4.65^{+0.26}_{-0.27}$\\
 \\
Lupus \\
\hline
  1 (IK~Lup) &     Sz 65  & 15h39m27.74s & -34d46m17.7s &      $110.00^{+0.10}_{-0.11}$ &     $61.41^{+0.10}_{-0.09}$ &     $-4.06^{+0.09}_{-0.09}$ &    $8.48^{+0.06}_{-0.06}$  \\
  2 (GW~Lup) &     Sz 71  & 15h46m44.70s & -34d30m36.3s &      $36.24^{+0.22}_{-0.22}$ &     $38.36^{+0.13}_{-0.14}$ &   $-1.96^{+0.18}_{-0.19}$ &    $2.49^{+0.18}_{-0.17}$\\
  3 &              2MASS J16124373-3815031 & 16h12m43.73s & -38d15m03.6s &      $17.34^{+0.78}_{-0.77}$ &   $51.83^{+0.63}_{-0.69}$ &     $-0.06^{+0.27}_{-0.29}$ &    $-1.15^{+0.30}_{-0.31}$ \\
  4 (HM~Lup) &     Sz 72 & 15h47m50.60s & -35d28m36.0s &      $20.00\pm0.70$* &     $31.50\pm0.60$* &     $-4.44^{+0.59}_{-0.55}$ &    $-2.66^{+0.52}_{-0.53}$\\
  5 &              Sz 77 & 15h51m46.93s & -35d56m44.7s &      $8.2\pm2.1$* &     $26.0\pm0.80$* &     $-0.5^{+1.8}_{-1.9}$ &    $1.1^{+1.8}_{-1.8}$ \\
  6 &             2MASS J16085324-3914401 & 16h08m53.22s & -39d14m40.7s &      $104.8^{+4.1}_{-4.1}$ &   $32.2^{+2.0}_{-2.2}$ &     $2.30^{+0.39}_{-0.40}$ & $-0.38^{+0.40}_{-0.39}$ \\
  7 &             Sz 131 & 16h00m49.41s & -41d30m04.4s &      $160.0^{+3.6}_{-3.6}$ &    $46.7^{+2.7}_{-2.7}$ &     $-0.6^{+1.1}_{-1.1}$ &    $3.2^{+1.2}_{-1.2}$ \\
  8 &              Sz 66 & 15h39m28.25s & -34d46m18.7s &      $83.1^{+1.9}_{-1.6}$ &    $76.8^{+5.4}_{-4.0}$ & $22.00^{+0.50}_{-0.48}$ &    $7.34^{+0.45}_{-0.47}$\\
  9 &              Sz 95 & 16h07m52.29s & -38d58m06.6s &      $159.5^{+6.9}_{-6.6}$ &  $54.64^{+6.1}_{-6.7}$ &   $-1.3^{+2.1}_{-2.1}$ &    $3.9^{+2.1}_{-2.2}$ \\
 10 &          V1094~Sco & 16h08m36.18s & -39d23m02.5s &      $107.87^{+0.06}_{-0.06}$ &   $52.02^{+0.04}_{-0.04}$ &     $-7.15^{+0.36}_{-0.36}$ &    $-3.83^{+0.26}_{-0.26}$\\
\\ 
Upper Scorpius \\
\hline
  1 &              2MASS J16120668-3010270 & 16h12m06.66s & -30d10m27.6s &      $45.10^{+0.20}_{-0.90}$\dag &     $37.00^{+0.10}_{-0.20}$\dag &     $4.6^{+20}_{-20}$\dag &    $6.6^{+10}_{-10}$\dag  \\
  2 &              2MASS J16054540-2023088 & 16h05m45.38s & -20d23m09.3s &      $49.4^{+2.7}_{-2.5}$ &     $56.4^{+2.6}_{-2.3}$ &   $-0.61^{+0.41}_{-0.41}$ &    $-0.85^{+0.39}_{-0.39}$ \\
  3 &              2MASS J16020757-2257467 & 16h02m07.56s & -22d57m47.4s  &   $78.1\pm9.6$* &     $58.20\pm0.13$* &    $-1.3^{+1.8}_{-1.7}$ &    $-0.2^{+1.2}_{-1.2}$ \\
  4 &              2MASS J16111742-1918285 & 16h11m17.41s & -19d18m29.2s &      $69^{+18}_{-11}$ &     $79.3^{+7.5}_{-18.8}$ &     $-19^{+36}_{-33}$ &    $0^{+20}_{-22}$\\
  5 (BV Sco) &     2MASS J16145026-2332397 & 16h14m50.25s & -23d32m40.2s &      $139.0\pm5.3$* &     $15\pm10$* &     $-2.0^{+3.5}_{-3.2}$ &    $-5.9^{+2.8}_{-2.9}$ \\
  6 &             2MASS J16163345-2521505 & 16h16m33.43s & -25d21m51.2s &      $59.8^{+2.5}_{-2.5}$ &   $62.9^{+2.6}_{-2.7}$ &     $2.3^{+7.2}_{-7.0}$ &    $-6.2^{+5.4}_{-5.4}$ \\
  7 &             2MASS J16202863-2442087 & 16h20m28.62s & -24d42m09.2s &      $179^{+11}_{-11}$\ddag &    $32.4^{+6.2}_{-9.0}$\ddag &     $0.6^{+7.0}_{-6.7}$\ddag &    $4.6^{+7.4}_{-7.5}$\ddag \\
  8 &              2MASS J16221532-2511349 & 16h22m15.32s & -25d11m35.7s &      $16.27^{+0.32}_{-0.33}$ &     $56.24^{+0.28}_{-0.29}$ &     $-1.92^{+0.23}_{-0.24}$ &    $-0.48^{+0.27}_{-0.27}$\\
  9 &              2MASS J16082324-1930009 & 16h08m23.25s & -19d30m01.0s &      $125.40^{+0.10}_{-0.09}$ &   $75.10^{+0.12}_{-0.13}$ &     $-1.61^{+0.25}_{-0.26}$ &    $0.55^{+0.19}_{-0.20}$ \\
 10 &          2MASS J16090075-1908526 & 16h09m00.74s & -19d08m53.3s &      $155.73^{+0.26}_{-0.26}$ &   $49.14^{+0.22}_{-0.21}$ &     $-3.13^{+0.17}_{-0.17}$ &    $0.77^{+0.18}_{-0.18}$\\ 
\enddata
\tablecomments{First two columns indicate the AGE-PRO naming convention (1 to 10 in each region) and the most common literature names of each source. Coordinates refer to the center of emission identified in the image plane for the continuum-only datasets of each source. Geometries and phase offsets were obtained through fitting azimuthally symmetric Gaussians to the dust visibilities. Disk geometries with an asterisk (*) come from fitting two-dimensional Gaussians to the \textsuperscript{12}CO moment zero maps (presented in \citealp{AGEPRO_XI_gas_disk_sizes}). \dag: Upper Scorpius 1 geometry and phase center  come from the work of \citet{2024ApJ...974..102S}. \ddag: Upper Scorpius 7 information refers to the quiescent data only (see \citealp{AGEPRO_XII_mm_var_USco7} for more information).
}
\end{deluxetable*}

\begin{deluxetable*}{llcccccccr}
\rotate
\tabletypesize{\footnotesize} 
\setlength{\tabcolsep}{2.5pt}
\tablecaption{Angular and Physical Radii Containing $68\%$, \textbf{$90\%$, and $95\%$} of the Continuum Emission of the AGE-PRO Disks (R\textsubscript{$68\%$}, R\textsubscript{$90\%$}\textbf{, and R\textsubscript{$95\%$}, }respectively) \label{Table2}}
\tablewidth{0.6\textwidth}
\tablehead{
\colhead{AGE-PRO ID} &
\colhead{Source} & \colhead{R\textsubscript{$95\%$}} &\colhead{R\textsubscript{$90\%$}} &
\colhead{R\textsubscript{$68\%$}} & \colhead{R\textsubscript{$95\%$}} &\colhead{R\textsubscript{$90\%$}} & \colhead{R\textsubscript{$68\%$}} & \colhead{Frank FWHM} & \colhead{Note} \\
\colhead{} & \colhead{} & \colhead{(arcsec)} &
\colhead{(arcsec)} & \colhead{(arcsec)} & 
\colhead{(au)} & \colhead{(au)} &
\colhead{(au)} & \colhead{(au)}
}
\startdata
Ophiuchus  \\
\hline
  1 (ISO-Oph 37) & SSTc2d J162623.6-242439   & $0.7506\pm{0.0010}$  &       $0.6866\pm{0.0020}$ &     $0.5038\pm{0.0018}$ & $104\pm{15}$  & $95\pm{14}$ &     $70\pm{10}$  & 11 & irregular structure  \\
  2 (ISO-Oph 94) & SSTc2d J162703.6-242005   &  $0.5391\pm{0.0019}$ &       $0.4997\pm{0.0008}$ &    $0.3768\pm{0.0005}$ & $75\pm{11}$ & $69\pm{10}$ &     $52.4\pm{7.6}$   & 15 &  irregular structure   \\
  3 (ISO-Oph 129) & SSTc2d J162719.2-242844   & $0.1409\pm{0.0039}$  &       $0.1256\pm{0.0023}$ &    $0.0918\pm{0.0011}$ & $19.6\pm{3.4}$ & $17.5\pm{2.9}$ &     $12.8\pm{2.0}$   & 10 &   \\
  4 (Elias 2-32) & SSTc2d J162728.4-242721   & $0.151\pm{0.060}$  &       $0.130\pm{0.027}$ &     $0.089\pm{0.016}$ & $21\pm{12}$  &$18.1\pm{6.9}$ &     $12.4\pm{4.3}$  & 18 &  tentative structure \\
  5 (ISO-Oph 161) & SSTc2d J162737.2-244237   & –  &      $<0.040$ &     $<0.026$ & – & $<5.5$ &     $<3.7$   &  12 & marginally resolved   \\
  6 (ISO-Oph 165) & SSTc2d J162738.9-244020   &  $0.3227\pm{0.0022}$ &       $0.28882\pm{0.00063}$ &     $0.21853\pm{0.00043}$ &  $44.9\pm{6.8}$  &   $40.1\pm{5.9}$ &     $30.4\pm{4.4}$    &   12 & irregular structure \\
  7 (IRS 63) & SSTc2d J163135.6-240129   & $0.6576\pm{0.0016}$  & $0.5294\pm{0.0007}$ & $0.3681\pm{0.0001}$ & $91\pm{13}$   & $74\pm{11}$ & $51.2\pm{7.4}$ &   8 & irregular structure \\
  8 & 2MASS J16230544-2302566   &  $0.289\pm{0.020}$ &       $0.248\pm{0.016}$ &     $0.169\pm{0.014}$ & $40.1\pm{8.9}$ & $34.5\pm{7.5}$ &     $23.5\pm{5.6}$   & 19 & irregular structure  \\
  9 (ISO-Oph 43) & SSTc2d J162627.5-244153   &  $0.2515\pm{0.0026}$ &     $0.2323\pm{0.0033}$ &     $0.1874\pm{0.0010}$ & $35.0\pm{5.4}$ & $32.3\pm{5.2}$ &     $26.0\pm{3.9}$   &   15 & inner dust cavity   \\
 10 (ISO-Oph 127) & SSTc2d J162718.4-243915   &  $0.4115\pm{0.0031}$ &       $0.3904\pm{0.0024}$ &     $0.3122\pm{0.0022}$ & $57.2\pm{8.7}$ &   $54.3\pm{8.2}$ &     $43.4\pm{6.6}$   &   13 & irregular structure   \\
 \\
Lupus \\
\hline
 1 (IK~Lup) &     Sz 65   &  $0.251\pm{0.014}$ &       $0.2229\pm{0.0053}$ &     $0.1631\pm{0.0074}$& $38.4\pm{2.4}$&  $34.10\pm{0.95}$ &     $25.0\pm{1.2}$  & 20 &   \\
  2 (GW~Lup) &     Sz 71   &  $0.690\pm{0.012}$  &       $0.5911\pm{0.0073}$ &     $0.3873\pm{0.0045}$  & $105.8\pm{2.1}$&   $91.1\pm{1.4}$ &     $59.67\pm{0.85}$  &  17 & irregular structure \\
  3 &              2MASS J16124373-3815031   & $0.260\pm{0.011}$  &      $0.1987\pm{0.0067}$ &     $0.1224\pm{0.0010}$  & $41.2\pm{1.8}$ &   $31.5\pm{1.2}$ & $19.42\pm{0.21}$  &  27 & \\
  4 (HM~Lup) &     Sz 72   &  – &       $<0.074$ &     $<0.049$ & – &  $<12$ &     $<7.7$  &  22 & marginally resolved   \\
  5 &              Sz 77   &  – &       $<0.18$ &     $<0.12$ &  – & $<28$ &     $<19$  &  28 & marginally resolved   \\
  6 &             2MASS J16085324-3914401   & $0.224\pm{0.076}$  &       $0.1347\pm{0.024}$ &     $0.0767\pm{0.0026}$ & $36.2\pm{12.7}$  & $21.8\pm{4.1}$ &     $12.39\pm{0.53}$  &   16 & tentative structure   \\
  7 &             Sz 131   & $0.224\pm{0.042}$  &      $0.191\pm{0.019}$ &     $0.1281\pm{0.0011}$ & $35.6\pm{6.9}$&  $30.4\pm{3.2}$ &     $20.39\pm{0.28}$  &   21 &  \\
  8 &              Sz 66   &  – &  $<0.13$ &     $<0.09$ & – & $<21$ &     $<14$  &  27 & marginally resolved   \\
  9 &              Sz 95   & –  &       $<0.19$ &     $<0.13$ & – & $<31$ &     $<21$  &  39 & marginally resolved    \\
 10 &          V1094~Sco   &  $2.7790\pm{0.0028}$ &       $2.2807\pm{0.0027}$ &     $1.44033\pm{0.00082}$ & $430.1\pm{2.6}$ & $353.0\pm{2.2}$ &     $222.9\pm{1.2}$  &  14 & irregular structure  \\
\\ 
Upper Scorpius \\
\hline
  1 &              2MASS J16120668-3010270   &  $0.7023\pm{0.0052}$ &       $0.6644\pm{0.0032}$ &     $0.6059\pm{0.0016}$ & $92.6\pm{0.91}$  &$87.64\pm{0.62}$ &     $79.92\pm{0.40}$  & 13 & inner dust cavity \\
  2 &              2MASS J16054540-2023088   &  – &       $<0.13$ &     $<0.084$ & – & $<17$ &     $<12$  & 21 & marginally resolved  \\
  3 &              2MASS J16020757-2257467   & $0.217\pm{0.061}$  &    $0.193\pm{0.047}$ &     $0.140\pm{0.013}$ &  $30.3\pm{8.7}$& $27.0\pm{6.6}$ &     $19.6\pm{1.9}$  & 31 &\\
  4 &              2MASS J16111742-1918285   &  – &       $<0.41$ &     $<0.27$ & – & $<56$ &     $<37$  & 53 & marginally resolved \\
 5 (BV Sco)  &     2MASS J16145026-2332397   & –  &      $<0.13$ &     $<0.087$ & – & $<19$ &     $<12$  & 20 & marginally resolved \\
  6 &             2MASS J16163345-2521505   &  $0.43\pm{0.19}$ &       $0.40\pm{0.12}$ &     $0.309\pm{0.033}$ & $69\pm{30}$ & $63\pm{19}$ &     $48.9\pm{5.4}$  & 37 & inner dust cavity\\
  7 &             2MASS J16202863-2442087   & $0.33\pm{0.12}$\ddag  &       $0.31\pm{0.14}$\ddag &     $0.2501\pm{0.0048}$\ddag &  $51\pm{18}$\ddag & $47\pm{22}$\ddag &     $38.19\pm{0.92}$\ddag   & 20 & inner dust cavity \\
  8 &              2MASS J16221532-2511349   & $0.2115\pm{0.0013}$ &       $0.19565\pm{0.00066}$ &     $0.15834\pm{0.00064}$ & $29.39\pm{0.27}$ & $27.19\pm{0.18}$ &     $22.01\pm{0.16}$  & 14 & inner dust cavity\\
  9 &              2MASS J16082324-1930009   & $0.3770\pm{0.0013}$  &       $0.349\pm{0.0014}$ &     $0.2768\pm{0.0012}$ &  $51.63\pm{0.34}$ &$47.80\pm{0.34}$ &     $37.91\pm{0.28}$  & 19 & irregular structure\\
 10 &          2MASS J16090075-1908526   &  $0.331\pm{0.019}$ &       $0.307\pm{0.010}$ &     $0.241\pm{0.014}$ & $45.4\pm{2.7}$ & $42.0\pm{1.5}$ &     $33.0\pm{2.0}$  & 12 & irregular structure \\ 
\enddata
\tablecomments{`Frank FWHM' describes the simulated \texttt{Frankenstein} resolution element for each source. First two columns indicate the AGE-PRO naming convention (1 to 10 in each region) and the most common literature names of each source. Lupus and Upper Scorpius radii in astronomical units are obtained by using the geometric Gaia DR3 distances of \citet{2021AJ....161..147B}. For Ophiuchus, an averaged distance of $139\pm20$ pc is assumed (\citealp{2021AA...652A...2G}). Upper limits are denoted with `$<$'. Upper Scorpius 7 (\ddag) information refers to the quiescent data only (see \citealp{AGEPRO_XII_mm_var_USco7} for more information). Last column describes the identified substructures with the categories presented in Sect. \ref{Frank_subsec}.}
\end{deluxetable*}

\section{Radial Profiles and Dust-disk Radii}\label{Frank}


\texttt{Frankenstein} is a Python library that fits nonparametrically the radial brightness profile of a disk using the observed visibilities and assuming azimuthal symmetry (\citealp{2020MNRAS.495.3209J}). We use \texttt{Frankenstein} because it allows for direct nonparametric fitting of the visibilities, enabling the recovery of subbeam features (compared to the CLEAN image) while exploiting the full sensitivity of the data.

The inputs of \texttt{Frankenstein} are the geometries and phase centers identified in Sect. \ref{sec:methodology} (Table \ref{Table1}). We run a grid of 17 different \texttt{Frankenstein} fits for each source (using the \textit{LogNormal} method), exploring the full range of geometry and phase center uncertainties, and the full range of \texttt{Frankenstein} hyperparameters recommended in \citet[$1.05 \le \alpha \le 1.3$ and $10^{-4} \le$ \textit{wsmooth} $\le 10^{-1}$]{2020MNRAS.495.3209J}.\footnote{The grid is [1.05, $10^{-2}$], [1.10, $10^{-2}$], [1.20, $10^{-2}$], [1.30, $10^{-2}$], [1.20, $10^{-1}$], [1.20, $10^{-3}$], [1.20, $10^{-4}$], [1.05, $10^{-4}$], [1.30, $10^{-1}$]. To this we add [1.20, $10^{-2}$] with $\textrm{PA}+\sigma{(\textrm{PA})}$, $\textrm{PA}-\sigma{(\textrm{PA})}$, $\textrm{inc}+\sigma{(\textrm{inc})}$, $\textrm{inc}-\sigma{(\textrm{inc})}$, $\Delta{\textrm{RA}}+\sigma{(\Delta{RA})}$, $\Delta{\textrm{RA}}-\sigma{(\Delta{RA})}$, $\Delta{\textrm{Dec}}+\sigma{(\Delta{Dec})}$, $\Delta{\textrm{Dec}}-\sigma{(\Delta{Dec})}$ for a total of 17 \texttt{Frankenstein} models.}

In some cases (Lupus 1, 3, 7, and Upper Scorpius 10),
the real component of the \texttt{Frankenstein} fit to the observed visibilities did not converge to zero brightness at the higher deprojected uv-distances. This could cause emission to be inferred on scales that are not well sampled in the uv-plane of our observations, specifically unresolved emission at short separation from the phase center. In these cases, we added a point source of the necessary flux to force the \texttt{Frankenstein} model to end at zero brightness. This point source was then removed to calculate the residuals of the model (Sect. \ref{residuals}). \textbf{Not surprisingly,} \texttt{Frankenstein} fails to produce a fit to the visibilities for the eight marginally resolved sources (Sect. \ref{galario}).

{The intrinsic angular resolution of the \texttt{Frankenstein} fit to each source is computed following the methodology described in \citet{2024ApJ...974..306S} and implemented in \citet{2024ApJ...971..129C}. The visibilities of a delta ring are sampled with the same uv-coverage of each observation, and then, they are fitted with \texttt{Frankenstein} using the same hyperparameters as for the original data. We locate the delta ring at R\textsubscript{max}/2, with R\textsubscript{max} being the maximum radius given to \texttt{Frankenstein} for fitting the original data. The obtained \texttt{Frankenstein} solution is similar to a Gaussian function (interpreted as a point-spread function), with an FWHM that depends on the uv-coverage. \textbf{We take this FWHM as the \texttt{Frankenstein} resolution element. In AGE-PRO data,} typical values for the \texttt{Frankenstein} resolution element are of $\sim0.09-0.15$", while the typical CLEAN beam size is of $\sim0.2-0.35$". Hence, on average, the \texttt{Frankenstein} resolution elements we obtain for the AGE-PRO disks are 1.5 to 2.5 times smaller than the CLEAN beam (in agreement with \citealp{2022MNRAS.509.2780J,2022MNRAS.514.6053J}). The \texttt{Frankenstein} resolution elements of the visibility fitting for each source are reported in Table \ref{Table2}.}

\textbf{We find mean resolution elements of 13, 22, and 22 au for the sources at Ophiuchus, Lupus, and Upper Scorpius, respectively. This is due to the fact that the signal-to-noise ratio (SNR) of Ophiuchus observations is much higher than for Lupus and Upper Scorpius observations (which are similar among themselves), such that longer baselines have sufficient SNR to significantly influence the visibility fit. Hence, for the \texttt{Frankenstein}-based substructure identification of this work, we have similar resolution for Lupus and Upper Scorpius, and around a factor of 2 better resolution for Ophiuchus.}

\begin{figure*}[!ht]
\includegraphics[width=\textwidth]{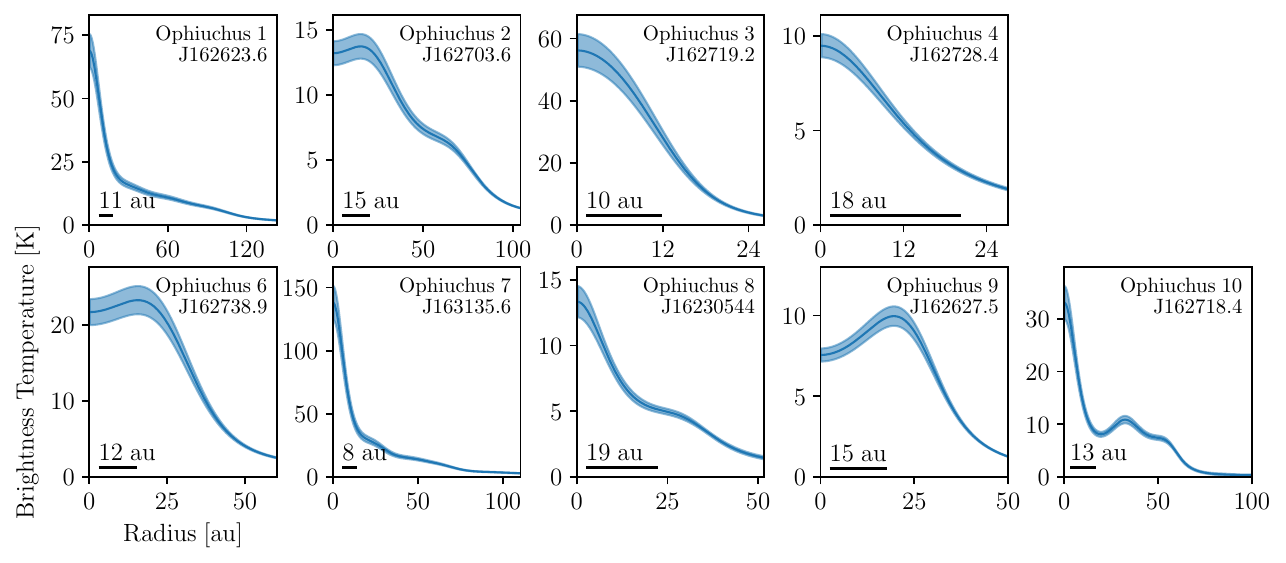}
\caption{Dust brightness temperature profiles of AGE-PRO Ophiuchus sources obtained from the best \texttt{Frankenstein} fit to each target and only for the resolved disks. The resolution element of the fit to the visibilities is shown at the bottom left corners (Table \ref{Table2}, Sect. \ref{Frank}). Plots show the radial profiles up to $1.5$R$_{90\%}$. Uncertainties are derived from the flux calibration error, \texttt{Frankenstein} error, and deviations from azimuthal symmetry.}\label{Plot: Oph}
\end{figure*} 

\begin{figure*}[ht!]
\includegraphics[width=\textwidth]{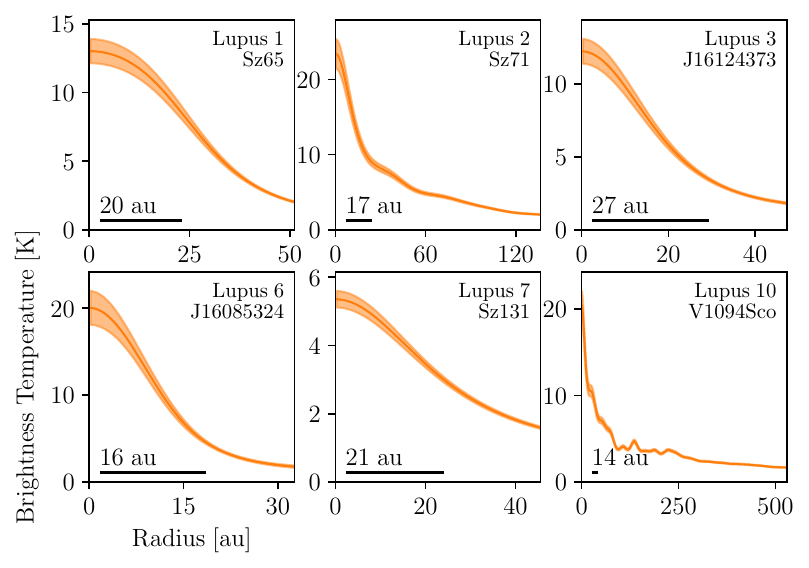}
\caption{Dust brightness temperature profiles of AGE-PRO Lupus sources obtained from the best \texttt{Frankenstein} fit to each target and only for the resolved disks. The resolution element of the fit to the visibilities is shown at the bottom left corners (Table \ref{Table2}, Sect. \ref{Frank}). Plots show the radial profiles up to $1.5$R$_{90\%}$. Uncertainties are derived from the flux calibration error, \texttt{Frankenstein} error, and deviations from azimuthal symmetry.}\label{Plot: Lupus}
\end{figure*} 

\begin{figure*}[ht!]
\includegraphics[width=\textwidth]{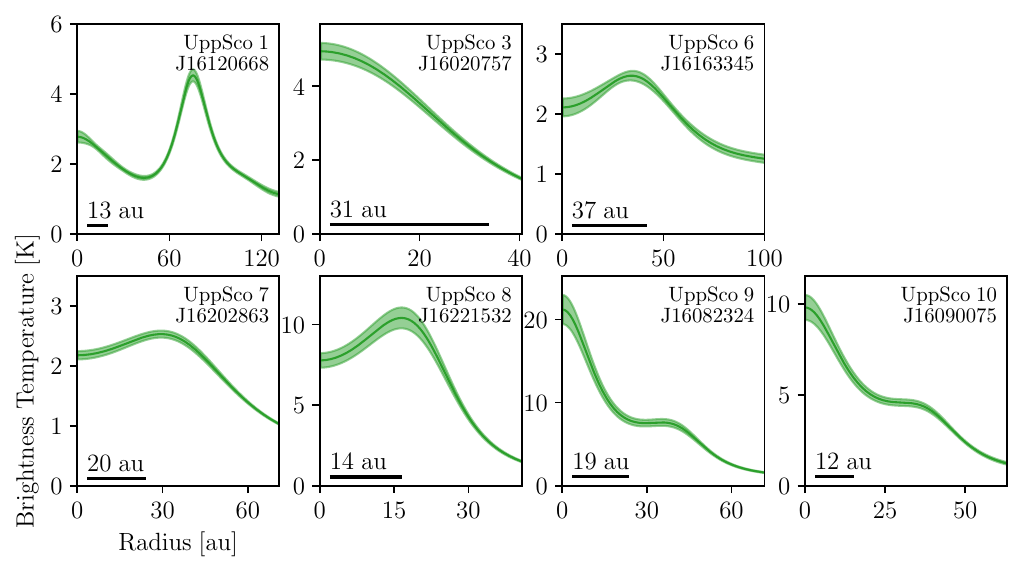}
\caption{Dust brightness temperature profiles of AGE-PRO Upper Scorpius sources obtained from the best \texttt{Frankenstein} fit to each target and only for the resolved disks. The resolution element of the fit to the visibilities is shown at the bottom left corners (Table \ref{Table2}, Sect. \ref{Frank}). Plots show the radial profiles up to $1.5$R$_{90\%}$. Uncertainties are derived from the flux calibration error, \texttt{Frankenstein} error, and deviations from azimuthal symmetry.}\label{Plot: UppSco}
\end{figure*}

\subsection{Dust-disk Radial Profiles}\label{Frank_subsec}


For each source, we visually inspected the fit to the visibilities of each \texttt{Frankenstein} solution and selected the best one using two criteria: minimal chi-squared between model and data, and convergence to zero brightness at the visibilities with higher deprojected uv-distances. If multiple fits provide similarly good solutions, we opted for the more flexible \texttt{Frankenstein} hyperparameters (lower $\alpha$ and \textit{wsmooth}, see \citealp{2020MNRAS.495.3209J}). The best \texttt{Frankenstein} brightness radial profile for each well-resolved AGE-PRO protoplanetary disk is presented in Fig. \ref{Plot: Three_regions} for all three regions. 

For 15 sources, substructures are apparent from the best \texttt{Frankenstein} fit. Seven sources show no substructure at the resolution of AGE-PRO (visibility fitting resolution element of $\sim0.09-0.15$"). The other eight sources are barely resolved in the continuum visibilities, and no radial profiles could be obtained for them (Sect. \ref{Frank_subsec}). Individual profiles are presented in Figs. \ref{Plot: Oph}, \ref{Plot: Lupus}, and \ref{Plot: UppSco} for Ophiuchus, Lupus, and Upper Scorpius sources, respectively. {Temperature brightness radial profiles are computed from the radial intensity profiles using the full blackbody equation (the Rayleigh-Jeans approximation is not assumed), and a central frequency of 230 GHz.} 



We group the {radial intensity profiles of the dust continuum emission} into three categories: \textit{inner dust cavity} group, for the disks with an inner dust cavity and a ring; \textit{irregular substructure} group, for the disks with a nonsmooth irregularly structured dust radial profile but with no resolved inner dust cavity; {and \textit{not-structured} group, if the dust radial profile is merely an unstructured decline}. There are 5 disks with inner dust cavities (Ophiuchus 9, and Upper Scorpius 1, 6, 7, 8) and 10 irregularly structured disks (Ophiuchus 1, 2, 6, 7, 8, 10, Lupus 2, 10, and Upper Scorpius 9, 10). The inner dust cavities are reminiscent of the so-called \textit{transition disks} (\citealp{1989AJ.....97.1451S, 2023EPJP..138..225V}). In this work, we refrain from calling them as such as we do not explore the correspondence of our sources with the original spectral energy distribution (SED)-based classification of transition disks. Table \ref{Table2} summarizes our results on the presence of dust substructures in the AGE-PRO sample of protoplanetary disks.

Ophiuchus 2 and 6 have top flat brightness profile distributions that could resemble an inner dust cavity (Fig. \ref{Plot: Oph}). However, the brightness contrast and size of these structures are too small to clearly discern them as cavities. We hence place Ophiuchus 2 and 6 in the irregular structure category.


There are seven not-structured disks at AGE-PRO resolution (Ophiuchus 3, 4, Lupus 1, 3, 6, 7, and Upper Scorpius 3). Of these later sources, we put down Ophiuchus 4 (Elias 2-32) and Lupus 6 (2MASS J16085324-3914401) as `tentatively substructured' as they show variations in their visibilities that might be indicative of substructure that \texttt{Frankenstein} was unable to model. 


In Fig. \ref{Plot: Mgas_vs_Rdust} we show how the type of substructure correlates with the gas-disk masses obtained in \citet{AGEPRO_V_gas_masses}. To limit the effect of resolution in the interpretation of this figure, we only show those sources for which our resolution is good enough that if substructures are present, we should have detected them in most cases (see Sect. \ref{Sec_sample_selection} for more information). We see that disks with large inner dust cavities have gas-disk masses compatible with the median gas-disk masses of their regions (\citealp{AGEPRO_V_gas_masses}). However, the structured disks without inner dust cavities (the \textit{irregular substructure} group) have gas-disk masses 1 to 2 orders of magnitude above the median value of their regions. This is further discussed in Sect. \ref{discuss_structures}.

We can evaluate the radial intensity profiles we present in this work by comparing with other observations. Among the AGE-PRO sample, there are seven sources with publicly available observations, which have angular resolutions higher than that of AGE-PRO. We compare these observations with \textbf{our radial intensity profiles }of the AGE-PRO dust continuum emission in Appendix \ref{pre_lit}. We find good agreement between the visibility modeled radial intensity profiles of this work (using the moderate resolution AGE-PRO observations) and the radial profiles obtained from observations at higher angular resolution. In some cases, there are substructures detected at high resolution that are not detected in the CLEAN images of AGE-PRO, but which are retrieved by the visibility modeled profiles of this work at the right locations. This comparison shows that we can trust the brightness radial profiles obtained in Sect. \ref{Frank_subsec} from visibility modeling for those sources in which there is no high angular resolution available. 


To summarize, we detect substructures in most of the well-resolved disks. Of the nine well-resolved Ophiuchus sources, seven show substructure (plus a tentative detection in Ophiuchus 4).  Of the six well-resolved Lupus sources, two show substructure (plus a tentative detection in Lupus 6).  Of the seven well-resolved Upper Scorpius sources, six show substructure. We provide a detailed discussion of the radial profile of each disk in Appendix \ref{final word}, and further discussion on the type and evolution of substructures in Sect. \ref{discuss_structures}.

\subsection{Dust-disk Radii}\label{sec_disk_radii}

The 17 \texttt{Frankenstein} brightness radial profiles \textbf{for each source} (Sect. \ref{Frank}) provide 17 different curves of growth, defined as the cumulative contribution to the source brightness as a function of the angular distance from the center. While there is no obvious way to properly define disk radii, in this work, we calculate the angular radii containing 68\%, 90\%\textbf{, and 95\%} of the emission (R\textsubscript{68\%}, R\textsubscript{90\%}\textbf{, and R\textsubscript{95\%}}, respectively). R\textsubscript{68\%} has often been identified as a better proxy for protoplanetary dust-disk size (\citealp{2023ASPC..534..501M}), although AGE-PRO sensitivity should be small enough to provide robust R\textsubscript{90\%} \textbf{and R\textsubscript{95\%}} values in most cases. The R\textsubscript{68\%}, R\textsubscript{90\%}\textbf{, and R\textsubscript{95\%}} radii for all sources appear tabulated in Table \ref{Table2}. Central values are the median from all 17 \texttt{Frankenstein} fits, and uncertainties are the maximum deviations from the median. Angular radii are converted to physical radii using the distance to each source (from Gaia DR3 \citealp{2021AJ....161..147B} geometric priors). AGE-PRO Ophiuchus sources do not have Gaia DR3 parallaxes, so we took the average distance to the cloud ($139\pm20$ pc, \citealp{2021AA...652A...2G}). 


Eight sources are marginally resolved in the continuum visibilities (Sect. \ref{galario}), and none of the \texttt{Frankenstein} fits for them converges adequately. For these sources, we take as upper limits for the R\textsubscript{68\%} and R\textsubscript{90\%} radii 2 times and 3 times the $\sigma$ of Sect. \ref{galario} Gaussian fit, respectively (Eq. \ref{Eq.1}, Table \ref{Table2}). \textbf{No R\textsubscript{95\%} radii are reported for these marginally resolved disks.}

{We compare the R\textsubscript{68\%} and R\textsubscript{90\%} dust-disk radii derived in this work with the dust-disk radii derived in \citet{2020A&A...633A.114S, 2021A&A...649A..19S} for the Lupus sources, and in \citet{2020ApJ...895..126H} for the Lupus and Upper Scorpius sources.  This comparison is shown in Fig. \ref{Plot: radii_compa}. Our R\textsubscript{68\%} and R\textsubscript{90\%} results and upper limits are in agreement with these previously reported radii, considering the differences in data sensitivity and methodology. The exception to this is Lupus 4, for which our R\textsubscript{68\%} upper limit is smaller than the R\textsubscript{68\%} reported in \citet{2021A&A...649A..19S}, although there is agreement between both works at R\textsubscript{90\%}. Lupus 7 and Lupus 10 are larger than previously reported, although our central values are within the uncertainties of \citet{2021A&A...649A..19S}. Hence, the R\textsubscript{68\%} and R\textsubscript{90\%} dust-disk radii obtained in this section for the AGE-PRO sample of disks and tabulated in Table \ref{Table2} are consistent with previous studies.}

\begin{figure}[t!]
\includegraphics[width=\columnwidth]{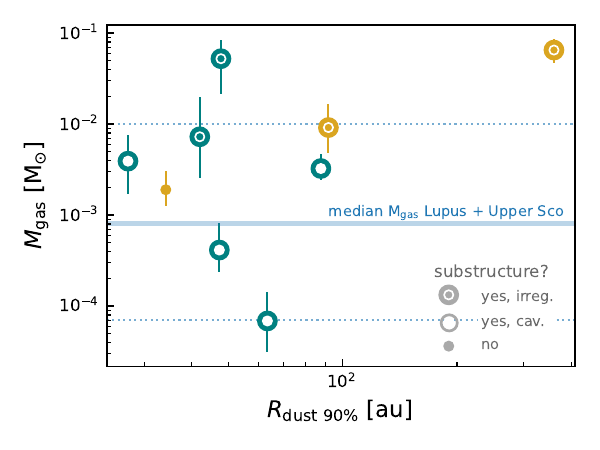}
\caption{Gas-disk mass from \citet{AGEPRO_V_gas_masses} vs. dust-disk R\textsubscript{90\%} radii only for those sources observed at enough resolution ($\theta_{D}/\theta_{res}>3$) to detect substructures in the majority of cases (see Sect. \ref{Sec_sample_selection}). Yellow markers indicate Lupus sources, and green markers Upper Scorpius sources. Lines show the median gas-disk mass in the full Lupus and Upper Scorpius AGE-PRO samples, and its 16$^{\rm th}$ and 84$^{\rm th}$ quantiles. Substructured disks without a large inner dust cavity (irregular) have abnormally high gas-disk masses, 1 to 2 orders of magnitude above the median value of their regions.}\label{Plot: Mgas_vs_Rdust}
\end{figure} 


\section{Nonazimuthally Symmetric Substructures}\label{residuals}




Once the best \texttt{Frankenstein} {radial profile of the dust continuum emission} is obtained for each disk (Sect. \ref{Frank_subsec}), we subtract the model visibilities from the data visibilities and then image the residuals. From the residuals, further nonazimuthally symmetric structure can be identified.


\begin{figure*}[t!]
\includegraphics[width=\textwidth]{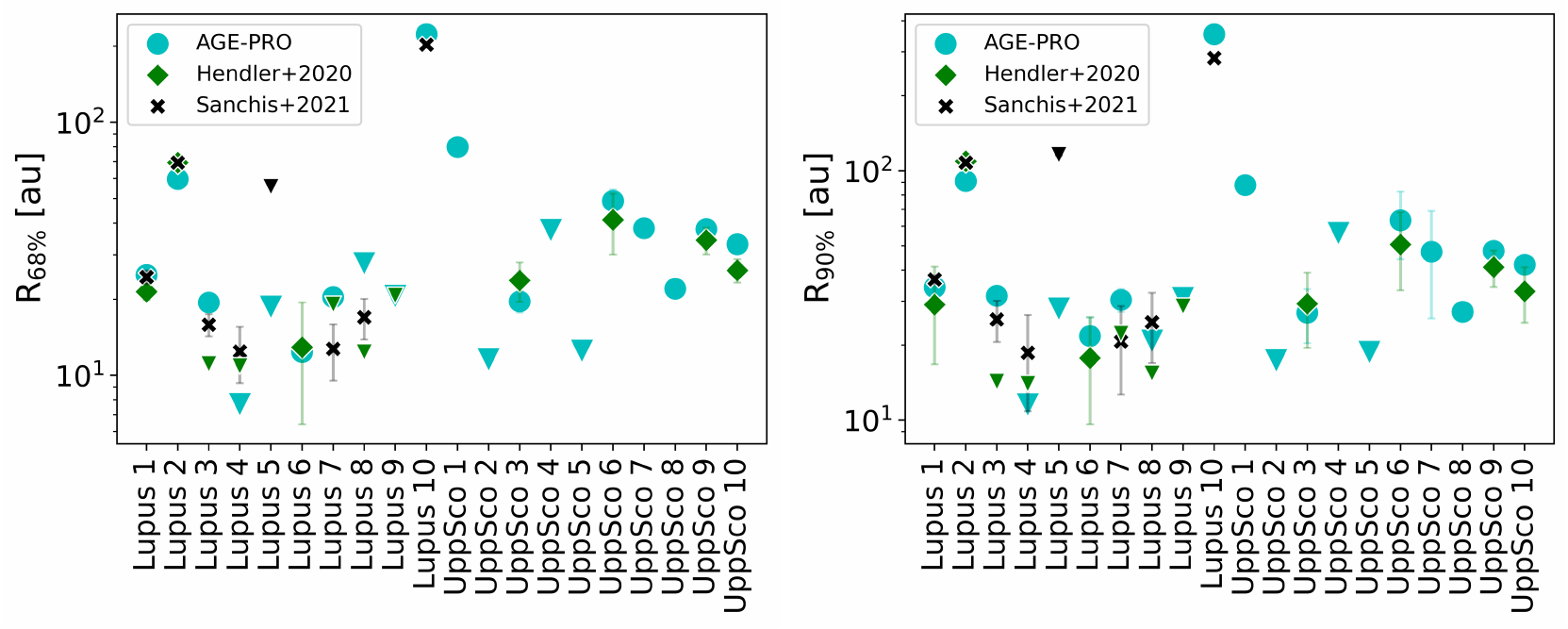}
\caption{Comparison between the dust-disk radii (R\textsubscript{68\%} and R\textsubscript{90\%}) obtained in this work and the dust-disk radii obtained in \citet{2020ApJ...895..126H} and \citet{2021A&A...649A..19S} for the AGE-PRO sources. Downward arrows denote upper limits.}\label{Plot: radii_compa}
\end{figure*}  




The obtained residuals are shown in Appendix \ref{final word} in Figs. \ref{Oph_residuals_mosaic}, \ref{Lupus_residuals_mosaic}, and \ref{UppSco_residuals_mosaic}, for Ophiuchus, Lupus, and Upper Scorpius sources, respectively. We only see residuals at the 5$\sigma$ level for one-third of the sample. \textbf{The well-resolved AGE-PRO sources with residuals at the 5$\sigma$ level are} Ophiuchus 1, 2, 7, 10, Lupus 2, 10, and Upper Scorpius 1, 8, and 10. The residuals of these sources are presented in Fig. \ref{Interest_residuals_mosaic}. All \textbf{of these sources} have detected substructures in their azimuthally symmetric radial profiles (Sect. \ref{Frank_subsec}). Residuals at the 3$\sigma$ level on the disk location are also found at the disks of Ophiuchus 6, Lupus 4, 9, and Upper Scorpius 3. We provide a detailed discussion for the residuals of each disk in Appendix \ref{final word}. 

\textbf{One relevant nonazimuthally symmetric structure identified by this analysis is in Ophiuchus 7 (IRS 63, SSTc2d J163135.6-240129), where a spiral is newly identified (Fig. \ref{Interest_residuals_mosaic}). \citet{2023AJ....166..184M} already found residuals highlighting nonaxisymmetric excess emission just south and north of the disk center. This source has been studied in detail (\citealp{2020Natur.586..228S,2021MNRAS.501.2934C,2023ApJ...958...98F,2023AJ....166..184M,2024A&A...688L..22P}). We believe this spiral was not detected before because it is a low-contrast structure that can be hidden at high resolution. Only the deeper AGE-PRO observations have allowed us to see the spiral (more information in Appendix \ref{final word}). \citet{2023AJ....166..184M} proposed this disk might be gravitationally unstable because of its high disk-to-star mass ratio (e.g. \citealp{2024Natur.633...58S}). \citet{2023ApJ...958...98F} further supported gravitational instability by quantifying the envelope-to-disk mass infall rate onto this source, which is much larger than the disk-to-star accretion rate. The detection of a spiral in the residuals of Fig. \ref{Interest_residuals_mosaic} is another piece evidence suggesting this disk is gravitationally unstable.}



\begin{figure*}[t!]
\includegraphics[width=\textwidth]{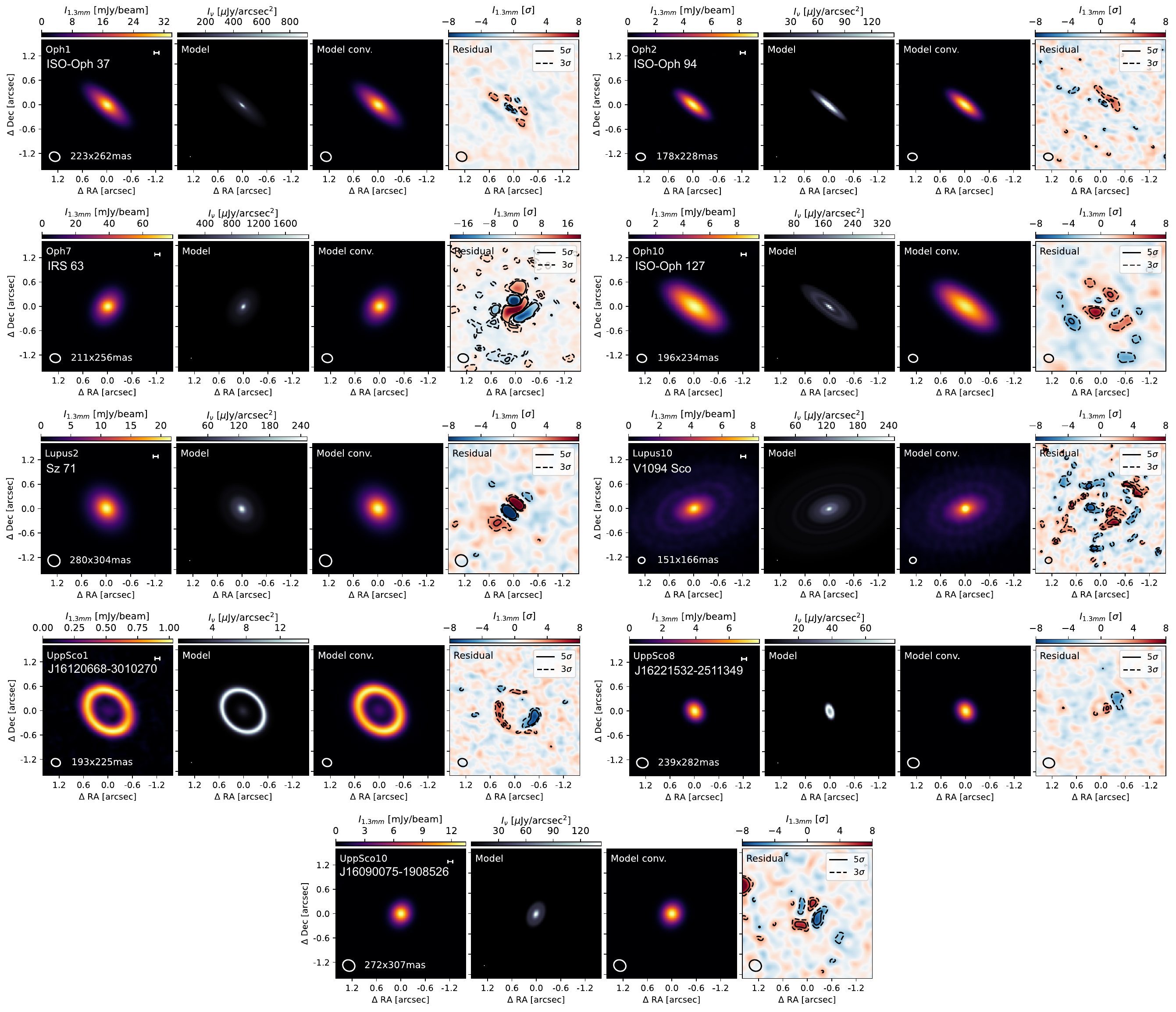}
\caption{\textbf{Mosaic of the well-resolved AGE-PRO sources with residuals }at the 5$\sigma$ level. For each source, we show the CLEAN image of the continuum-only data, the image of the model with pixel size, the image of the model convolved with the CLEAN beam, and the residual image from subtracting the visibilities of the data from those of the model. Residuals are shown in units of the rms noise value.}\label{Interest_residuals_mosaic}
\end{figure*}

\section{Analysis and Discussion}\label{analysis}

\begin{figure*}[t!]
\includegraphics[width=\textwidth]{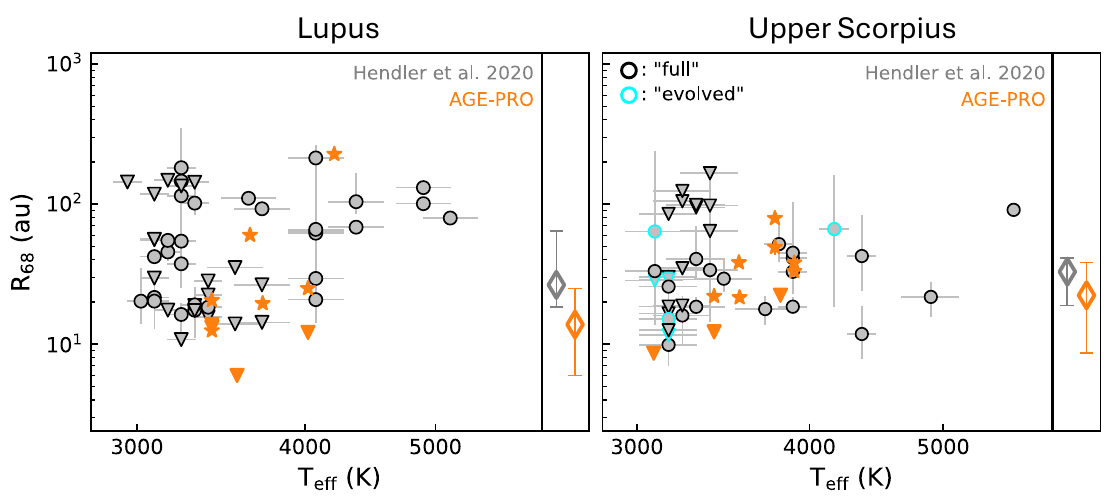}
\caption{Comparison of the Lupus and Upper Scorpius R\textsubscript{68\%} millimeter continuum radii from AGE-PRO (in orange) and the samples presented in \citet[in gray]{2020ApJ...895..126H}. Edge colors of the latter in Upper Scorpius are set to black or cyan depending on whether the disk is identified as `full' or `evolved' (following the classification of \citealp{2022AJ....163...25L}). For reference, all AGE-PRO sources are `full' disks but Upper Scorpius 4 (which is `transitional'). Downward triangles signify upper limits. The two open diamonds on the right of each plot are the median size computed using a Kaplar–Meier estimator that includes the effect of the upper limits, with the error bar indicating the 95\% confidence interval. These median sizes are limited to AGE-PRO spectral type range (K6–M3).}\label{Plot: Hendler_comp}
\end{figure*} 

In this section, we discuss the AGE-PRO sample of dust-disks from a population perspective, and show the general trends that arise between the Ophiuchus ($0.5-1$ Myr) sources, the Lupus ($1$-$3$ Myr) sources, and the Upper Scorpius ($2$-$6$ Myr) sources (for more details about these ages, see \citealp{AGEPRO_I_overview}). 

\textbf{\subsection{Sample selection}\label{Sec_sample_selection}}

We first put the AGE-PRO sample into perspective with respect to the three different star forming regions. In Fig. \ref{Plot: Hendler_comp} we compare the R\textsubscript{68\%} dust-disk radii of the \citet{{2020ApJ...895..126H}} compilations for Lupus and Upper Scorpius with the AGE-PRO samples of those regions. For deriving the median dust-disk sizes, we only consider the spectral type range of the AGE-PRO sample (K6–M3, \citealp{AGEPRO_I_overview}) to minimize the effect of underlying trends with stellar mass. For Upper Scorpius, the median R\textsubscript{68\%} of the \citet{{2020ApJ...895..126H}} sample and the median R\textsubscript{68\%} of the AGE-PRO sample are consistent ($32.9^{+8.3}_{-14}$ vs $22^{+16}_{-14}$ au, respectively). For Lupus, we find a somewhat smaller R\textsubscript{68\%} median value for the AGE-PRO sample ($13.8^{+11}_{-7.8}$ au vs. $26.5^{+38}_{-8.1}$ au in \citealp{2020ApJ...895..126H}), although they are consistent within error bars. However, the Lupus \textbf{analysis in \citet{{2020ApJ...895..126H}} was based on low-sensitivity ALMA data (\citealp{2016ApJ...828...46A})}, hence biasing the size measurements toward the brighter and thus larger disks of the Lupus population. This is confirmed by \citet{2025A&A...696A.232G}, who in a more sensitive Lupus survey find a median R\textsubscript{68\%} value for Lupus of $13.5$ au, in line with the AGE-PRO results (Fig. \ref{Plot: Hendler_comp}). We hence conclude that AGE-PRO is representative in\textbf{ dust-disk }radii of both Lupus and Upper Scorpius regions. 

It is harder to make a fair comparison of the AGE-PRO sample with the whole Ophiuchus region. Existing Ophiuchus surveys are shallow and AGE-PRO only selected Class Is from this region (\citealp{AGEPRO_I_overview}), which also contains a rich population of Class II sources (\citealp{2019MNRAS.482..698C}). Our AGE-PRO Ophiuchus sample has a median R\textsubscript{68\%} of $30$ au with a standard deviation of $20$ au. Comparing with \citet{2025ApJ...981L...4D}, it seems our sample is typical of the region in dust-disk size, although it leans toward large sizes.

\citet{AGEPRO_II_Ophiuchus}, \citet{AGEPRO_III_Lupus}, and \citet{AGEPRO_IV_UpperSco} papers have shown that the AGE-PRO Ophiuchus, Lupus, and Upper Scorpius samples are also not biased in millimeter or \textsuperscript{12}CO flux with respect to the entire K6–M3 population of their respective regions. We can hence conclude that, while small, the AGE-PRO sample of protoplanetary disks is not biased by its observational properties and is reasonably representative of the full K6–M3 disk population in the three considered regions. 



\textbf{Regarding the representativeness of the disk-substructures found in this work}, \citet{2023ASPC..534..423B} showed that if significant high resolution and sensitivity are used, substructures are detected in almost all disks. In particular, from a compilation of 479 young stars, if $\theta_{D}/\theta_{res}>10$, 95\% of the sources have substructures (with $\theta_{D}$ being the angular diameter containing 90\% of the emission, and $\theta_{res}$ the angular resolution of the observation). This percentage goes down to 50\% if $3<\theta_{D}/\theta_{res}<10$, and to 2\% if $\theta_{D}/\theta_{res}<3$. We compare these percentages with the AGE-PRO sample, taking as $\theta_{res}$ the \texttt{Frankenstein} resolution elements derived in Sect. \ref{Frank}. Of the AGE-PRO disks, Ophiuchus 1, 7, 10, Lupus 2, 10, and Upper Scorpius 1 have $\theta_{D}/\theta_{res}>10$. Of these six sources, we find substructures in all of them. Ophiuchus 2, 3, 6, 8, 9, Lupus 1, and Upper Scorpius 6, 7, 8, 9, 10 have $3<\theta_{D}/\theta_{res}<10$. Of these 11 sources, 9 have substructures (82\%), with the exceptions of Ophiuchus 3 (ISO-Oph 129) and Lupus 1 (Sz 65). The remaining 13 AGE-PRO sources have $\theta_{D}/\theta_{res}<3$. Of the latter, none show substructures (although we report Ophiuchus 4 and Lupus 6 as tentative detections in Sect. \ref{Frank_subsec}).  Comparing to \citet{2023ASPC..534..423B} results, our higher detection fraction in the intermediate resolution regime ($3<\theta_{D}/\theta_{res}<10$, 82\% vs. 50\%) could be due to the much higher than average sensitivity of the AGE-PRO program, or to the differences between detecting substructures from the visibility fitting instead of from the image plane. Hence, the fraction of substructures we have identified in this work considering the resolution we obtained agrees well with \citet{2023ASPC..534..423B} statistics on a larger population.

\vspace{13pt}
\subsection{Evolution of Dust-disk Radii with Age and across Regions}\label{Sec_population}

In Fig. \ref{Plot: FLux_Radii} we investigate the millimeter continuum size–luminosity relationship in the AGE-PRO sample. This correlation (first identified in \citealp{2010ApJ...723.1241A}) holds important information about the mechanisms driving the evolution of protoplanetary disks and the properties of dust particles in disks (e.g. \citealp{2017ApJ...845...44T, 2020MNRAS.492.1279S,2022MNRAS.514.1088Z}). In particular, \citet{2020ApJ...895..126H} showed that this relationship is not universal between regions. As disk size ($R$), we take the R\textsubscript{68\%} and R\textsubscript{90\%} radii derived in Sect. \ref{sec_disk_radii}. Following the procedure of \citet{2018ApJ...865..157A}, the continuum luminosities (L\textsubscript{mm}) are cast here in units of Jy for a distance of 140 pc, to ease comparison with other data sets. We fit power laws ($R=10^{\alpha}\cdot{}\text{L\textsubscript{mm}}^{\beta}$) to the whole AGE-PRO sample and to each separate star forming region using a Bayesian linear regression analysis that accounts for upper limits in the data (\citealp{1986ApJ...306..490I,2007ApJ...665.1489K}). We get the following results. For the Ophiuchus sources,

\begin{equation}
    R\textsubscript{68\%},\;\alpha=2.12\pm{0.40},\;\beta=0.42\pm{0.22},\;\rho=0.65.
\end{equation}
\begin{equation}        R\textsubscript{90\%},\;\alpha=2.23\pm{0.38},\;\beta=0.40\pm{0.20},\;\rho=0.66.
\end{equation}

For the Lupus sources,

\begin{equation}
    R\textsubscript{68\%},\;\alpha=2.26\pm{0.36},\;\beta=0.46\pm{0.18},\;\rho=0.73.
\end{equation}
\begin{equation}        R\textsubscript{90\%},\;\alpha=2.46\pm{0.36},\;\beta=0.47\pm{0.18},\;\rho=0.74.
\end{equation}

For all AGE-PRO sources, 

\begin{equation}
    R\textsubscript{68\%},\;\alpha=1.88\pm{0.18},\;\beta=0.23\pm{0.08},\;\rho=0.49.
\end{equation}
\begin{equation}    R\textsubscript{90\%},\;\alpha=2.02\pm{0.18},\;\beta=0.22\pm{0.08},\;\rho=0.51.
\end{equation}

These are with $\rho$ being the mean Pearson correlation coefficient of the Bayesian posterior. For the Upper Scorpius sources, there is no significant correlation, we find $\rho=0.18$ and $\rho=0.08$ for R\textsubscript{68\%} and R\textsubscript{90\%}, respectively. Uncertainties in the fit parameters indicate the standard deviations in the posterior distributions.\textbf{ As indicated by the low $\rho$ correlation coefficients and large uncertainties, the correlations we find are limited by the small AGE-PRO sample size.}




We find very similar correlations for R\textsubscript{68\%} and R\textsubscript{90\%}. For the Lupus sources, we find a correlation of $\text{L\textsubscript{mm}}\propto R^{2.2}$. This is in agreement with \citet{2020ApJ...895..126H} results. We also find a correlation for our Ophiuchus sample ($\text{L\textsubscript{mm}}\propto R^{2.4}$) compatible, yet steeper, than the one found by \citet{2020ApJ...895..126H} \textbf{for Class II sources of the region. In turn,} the central value we obtain for the Ophiuchus power law provides a flatter relation than the one found by \citet{2020ApJ...890..130T} in Class 0 and Class I protostars in Orion (L\textsubscript{mm}$\sim R^{3.3}$). This suggests that the Ophiuchus AGE-PRO sample (selected to be Class Is) is evolutionary somewhere between the Class 0/early Class I stage and the Class II stage. Regarding Upper Scorpius, if we restrict \citet{2020ApJ...895..126H} Upper Scorpius sample to the AGE-PRO stellar mass range, a Kendall Tau test on the detections also results in no R\textsubscript{dust}–L\textsubscript{mm} power-law correlation. This is not the case for Ophiuchus and Lupus \citet{2020ApJ...895..126H} samples, for which we still recover a correlation if restricting by stellar mass. This puts our results in agreement with \citet{2020ApJ...895..126H}. When the complete Upper Scorpius region is considered, Pinilla et al. (in prep.) does recover a flatter correlation for Upper Scorpius than for younger regions in the stellar mass range of AGE-PRO.

\begin{figure*}[t!]
\includegraphics[width=\textwidth]{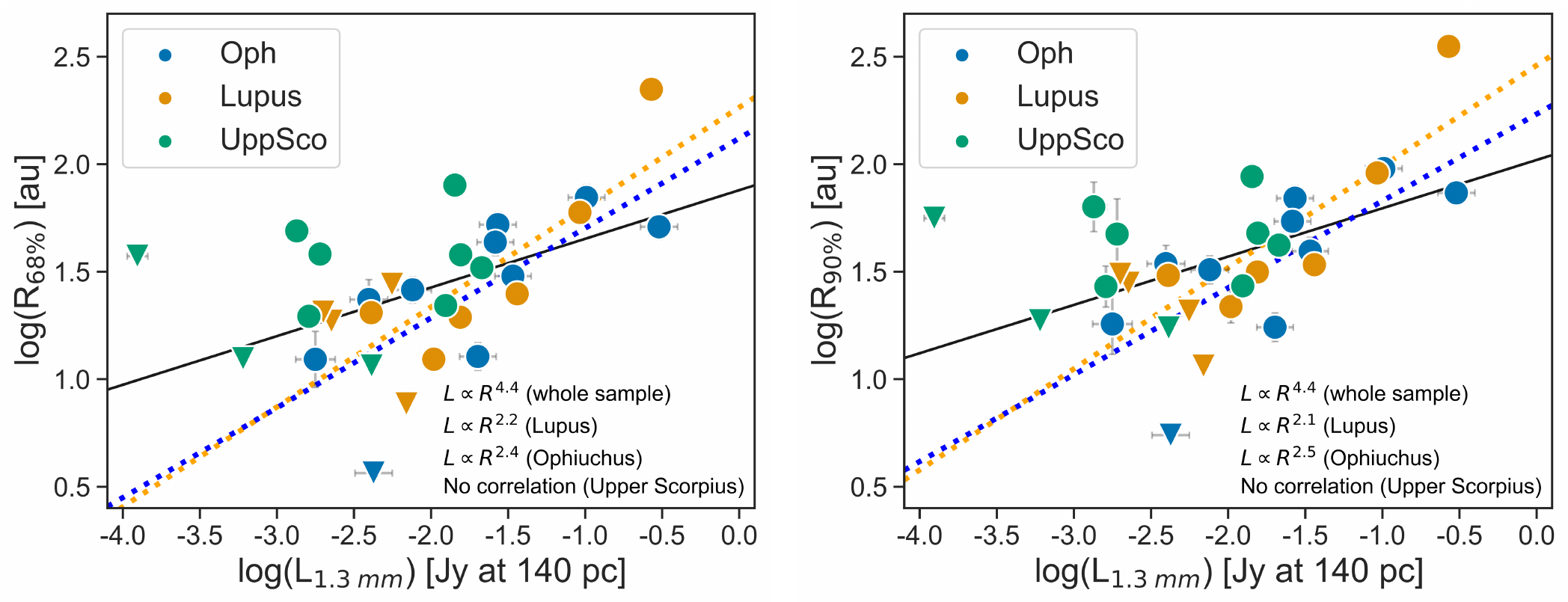}
\caption{Millimeter continuum size–luminosity relationship for R\textsubscript{68\%} and R\textsubscript{90\%} dust-disk sizes in the AGE-PRO sample. Black line is the fit to all AGE-PRO sources. The colored lines trace the best fit per region. Fits are derived from a Bayesian linear regression analysis that accounts for upper limits in the data. Upper Scorpius shows no significant correlation. Resulting power laws of every fit are indicated in each plot.}\label{Plot: FLux_Radii}
\end{figure*} 



The absence of a R\textsubscript{dust}–L\textsubscript{mm} significant correlation in the AGE-PRO Upper Scorpius sample can be interpreted in terms of dust evolution. \citet{2019MNRAS.486L..63R} proved that the R\textsubscript{dust}–L\textsubscript{mm} relation emerges naturally in the drift-dominated regime of dust grain evolution. However, if pressure dust traps are developed, the disk radii is fixed by the outer dust trap (\citealp{2022A&A...668A.104S,AGEPRO_VI_DustEvolution}). \textbf{Dust traps fixing the dust-disk radii in the Upper Scorpius sample also explain the observed R\textsubscript{95\%}/R\textsubscript{90\%} ratios, of order $\sim1.08$ for AGE-PRO Upper Scorpius sources, but of order $\sim1.12-1.31$ for the Lupus sources (with Lupus 6 having R\textsubscript{95\%}/R\textsubscript{90\%}} = 1.7)}. Indeed, \citet{2020A&A...635A.105P} and \citet{2022A&A...661A..66Z} predicted a flatter R\textsubscript{dust}–L\textsubscript{mm} correlation for disks with inner dust cavities, which are at least 40\% of our Upper Scorpius population (Sect. \ref{Frank_subsec}, Fig. \ref{Plot: UppSco}). We hence suggest that in Upper Scorpius there is a higher incidence of dust traps setting the dust-disk radii and flattening the R\textsubscript{dust}–L\textsubscript{mm} relation. This results in no significant correlation appearing in the case of AGE-PRO small sample size.







\begin{figure*}[t!]
\includegraphics[width=\textwidth]{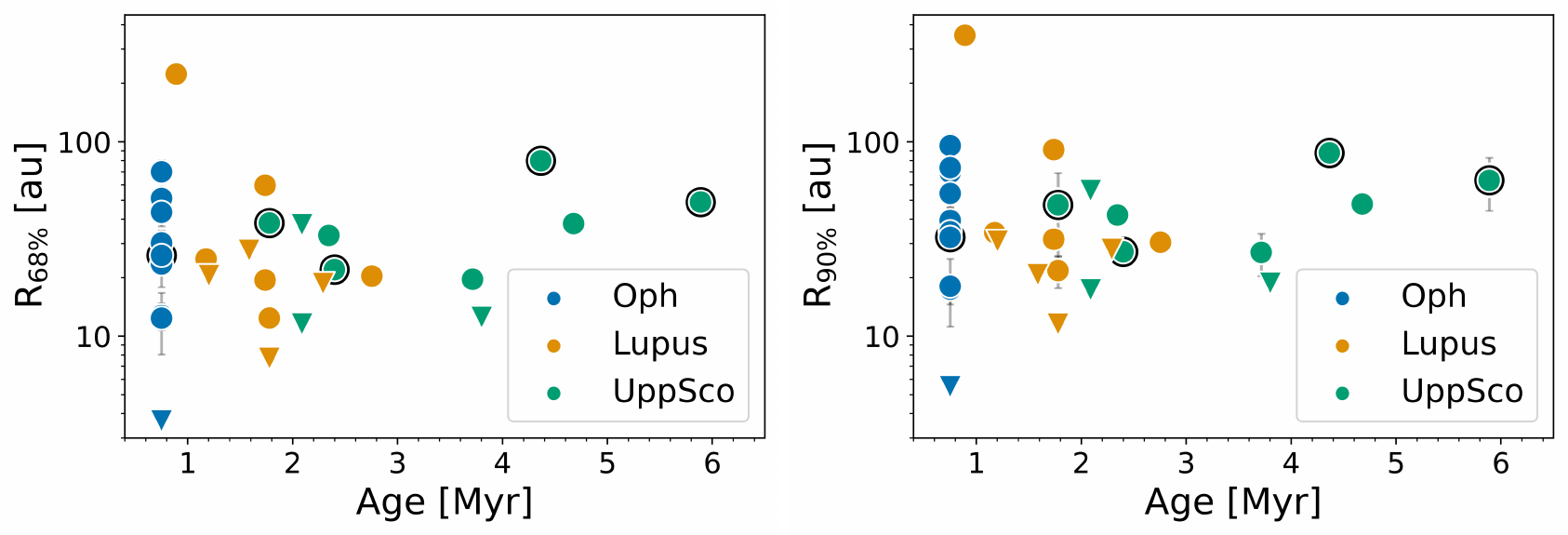}
\caption{R\textsubscript{68\%} and R\textsubscript{90\%} dust-disk radii vs. individual age for each AGE-PRO source. Disks with large inner dust cavities are highlighted with a black contour. Downward arrows denote upper limits. For uncertainties on the age determination, see \citet{AGEPRO_I_overview}.}\label{Plot: radii_compa_2}
\end{figure*}

In Fig. \ref{Plot: radii_compa_2} we show R\textsubscript{68\%} and R\textsubscript{90\%} dust-disk radii as a function of age for the AGE-PRO sample (for more details about these ages, see \citealp{AGEPRO_I_overview}). The absence of any trend with age is clear, as it is the absence of any evolution of the dust-disk radii across the three regions. \citet{2019MNRAS.486.4829R} predicted that, in smooth disks, the dust-disk R\textsubscript{68\%} decreases with time, as R\textsubscript{68\%} traces the grains that are large enough to have a significant submillimeter opacity. If there are substructures, this happens slower (\citealp{2020A&A...635A.105P, 2022A&A...661A..66Z, 2024A&A...688A..81D}). Hence, a higher incidence of dust traps at later ages also explains the apparent absence of dust-disk radii evolution with age in the AGE-PRO sample.

\textbf{We note that this is different from claiming an absence of dust-disk radii evolution with time. The AGE-PRO sample is only representative of the surviving population in Upper Scorpius. Lupus has a Class II disk fraction of $\sim50\%$ while the older Upper Scorpius region has a disk fraction of $\sim15\%$ (\citealp{2014A&A...561A..54R}). In addition, Fig. \ref{Plot: Hendler_comp} shows that the complete Upper Scorpius population has no sources with R\textsubscript{68\%}$>100$ au, while the Lupus region has several. This suggests an evolution toward smaller dust-disk sizes and disk-dissipation (i.e. \citealp{2025ApJ...978..117C}). As a result, the surviving disk population in Upper Scorpius has a higher fraction of disks with dust traps, which have prevented the inward drift and dissipation of their dust-disks.
}

\textbf{This conclusion connects with the results of other AGE-PRO papers. }\citet{AGEPRO_VI_DustEvolution} independently suggests that the dust content of the AGE-PRO disks in Upper Scorpius has mostly survived until their current age due to the presence of dust traps. Another support to this idea comes from the absence of evolution \textbf{in the AGE-PRO sample} of the $R_{\rm CO,\ 90\%}/R_{\rm dust,\ 90\%}$ ratio from Lupus to Upper Scorpius (\citealp{AGEPRO_XI_gas_disk_sizes}), which cannot be explained with smooth disks (\citealp{AGEPRO_VIII_ext_photoevap}).






We hence conclude that a high fraction of disks in Upper Scorpius with dust trapping substructures is the best explanation for the apparent lack of evolution of dust-disk radii with age \textbf{in the AGE-PRO sample} (Fig. \ref{Plot: radii_compa_2}), and the absence of significant R\textsubscript{dust}–L\textsubscript{mm} correlation in Upper Scorpius with respect to Ophiuchus and Lupus (these later following L\textsubscript{mm}$\sim R^{2}$, Fig. \ref{Plot: FLux_Radii}). We delve into this idea in the next section (Sect. \ref{discuss_structures}), where the detected dust substructures are discussed.




\subsection{The Evolution of Dust Substructures}\label{discuss_structures}

There are different possibilities for the origin of dust substructures, and their evolution mechanisms are not well understood (\citealp{2020ARA&A..58..483A,2023ASPC..534..423B,2023ASPC..534..605B, 2023ASPC..534..717D} and references therein). It seems  clear that substructures become easier to trace over time, as dust trapping acts, and we can achieve higher contrasts as the emission turns optically thinner (e.g. \citealp{2023ApJ...951....8O}). However, disk substructures are theorized to be ubiquitous and to appear early (\citealp{2019ApJ...872..112V,2020ApJ...902..141S,2024A&A...688A..81D}).




The main trend of dust substructures with age that can be appreciated in AGE-PRO is the much higher fraction of disks with large inner dust cavities ($>15$ au) in Upper Scorpius compared to Ophiuchus and Lupus (Fig. \ref{Plot: Three_regions}). Considering only those sources for which we have enough resolution ($\theta_{D}/\theta_{res}>3$) to detect substructures in the majority of cases (Sect. \ref{Sec_sample_selection}); four out of six disks in Upper Scorpius show inner dust cavities, for one out of eight in Ophiuchus (and this one, Ophiuchus 9, is among the most evolved sources of the Ophiuchus sample, \citealp{AGEPRO_II_Ophiuchus}), and zero out of three in Lupus. In contrast, six of the eight Ophiuchus sources show irregular structure, for two in both Lupus and Upper Scorpius. This suggests an evolution of \textbf{disk populations from having a high fraction of disks with} irregular substructures toward having \textbf{a high fraction of disks with} large inner dust cavities and a ring. 



In the AGE-PRO Lupus sample, 7 out of the 10 sources have $\theta_{D}/\theta_{res}<3$, which might have caused the fact that we are not effectively resolving substructures in Lupus. However, we note that cavities are often detected in Lupus (\citealp{2025A&A...696A.232G}) and the similar age region of Taurus (\citealp{2018ApJ...869...17L, 2024ApJ...966...59S}), \textbf{indicating that small unresolved cavities might be present in the AGE-PRO Lupus sample}. Hence, we claim that the transition toward disk populations with a high fraction of disks with large inner dust cavities happens at the $\sim 1-2$ Myr timescale (\textbf{Lupus age, Fig. \ref{Plot: radii_compa_2}}). 




Another piece of evidence favoring the evolution with time toward a larger fraction of disks with large inner dust cavities is presented in Fig. \ref{Plot: Mgas_vs_Rdust}. Fig. \ref{Plot: Mgas_vs_Rdust} shows that disks with large inner dust cavities have a gas-disk mass compatible with the median gas-disk mass of their regions (\citealp{AGEPRO_V_gas_masses}). This means that, at the age of $\sim 1-2$ Myr, most disks have regulated the evolution of their gas-disk mass in a similar way that disks with large inner dust cavities do (\citealp{2015A&A...579A.106V,2016A&A...585A..58V}), suggesting they also have developed similar dust traps. A different argumentation could be that disks with irregular structure but no large dust cavity have abnormally high gas-disk masses for their regions. These high gas-disk masses could be explained via late infall material and interaction with the environment (\citealp{2023EPJP..138..272K,2023ASPC..534..233P,2024ApJ...972L...9W,2024A&A...691A.169W}). This interaction could also be responsible for the irregular structure, as large scale structures at the outer edges of disks can be replicated easily within a relatively short period of infall (\citealp{2022ApJ...928...92K}). 



Large inner dust cavities are unlikely to have been created via photoevaporation alone (see \citealp{2023MNRAS.523.3318P, 2023EPJP..138..225V, 2024ApJ...974..306S}, and references therein), with planetary companions clearing a gap as the most promising explanation \textbf{(e.g. PDS 70, \citealp{2018A&A...617A..44K,2019NatAs...3..749H})}. \citet{2023A&A...679A..15G} and \citet{2024A&A...691A.155H} have shown that not very massive planets are needed to produce these inner cavities. \textbf{Indeed, there }are two AGE-PRO sources with inner dust cavities for which disk-to-star accretion rate measurements are available (\citealp{AGEPRO_I_overview}), and these cannot be reconciled with photoevaporating disks (\textbf{Fig. 6 of} \citealp{2017RSOS....470114E}). 


We hence propose an evolutionary sequence in which protoplanets at inner orbits have time to open cavities in $\sim 1-2$ Myr. This \textbf{would result in} disk populations \textbf{with} a higher fraction of large inner dust cavities at later ages. In turn, this fixes the outer dust-disk radii in a fraction of the population for which R\textsubscript{68\%} does not evolve over time and causes the flattening of the R\textsubscript{dust}-L\textsubscript{mm} relationship in Upper Scorpius (Sect. \ref{Sec_population}). \textbf{Disks that do not develop these dust traps keep reducing their dust-disk radii and are mostly dispersed or undetected at the Upper Scorpius age}. This evolutionary scenario would imply a \textbf{large fraction of protoplanetary disks with inner planets capable of creating large dust cavities}. This fraction is at least 40\% \textbf{of the disk-hosting Upper Scorpius population}, based on our limited sample size statistics. This large fraction can be reconciled with the exoplanet demographics observed around main-sequence stars (e.g. \citealp{2024arXiv240916993K}). These planets must have formed fast (within $\sim 2$ Myr), which is compatible with theory (e.g. \citealp{2024A&A...688A..22L}). As a final piece of argumentation, we already have a tentative detection for a massive inner protoplanet in the dust cavity of one of the AGE-PRO disks (Upper Scorpius 1, \citealp{2024ApJ...974..102S}).

Similar evolutionary sequences have been previously suggested from observational evidence (e.g. \citealp{2018A&A...620A..94G, 2020A&A...633A..82G, 2021MNRAS.501.2934C, 2021AJ....162...28V, 2022MNRAS.511.2453P,2025ApJ...984L..57O}). This AGE-PRO work provides the first evidence for an evolution of protoplanetary disk populations toward having higher fractions of \textit{transition disks} based on a representative sample of sources in three different star forming regions. 


\begin{figure*}[t!]
\includegraphics[width=\textwidth]{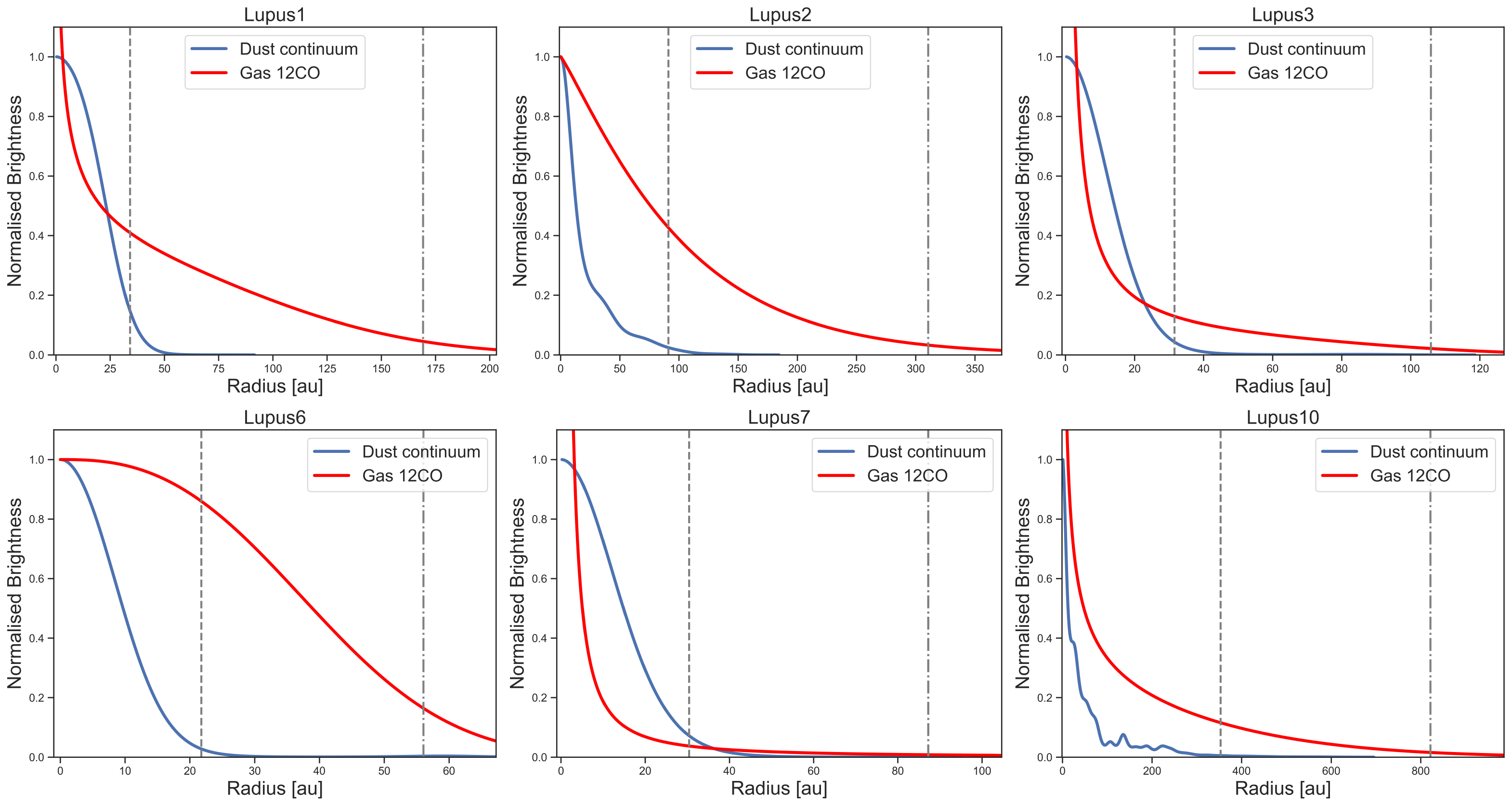}\\
\includegraphics[width=\textwidth]{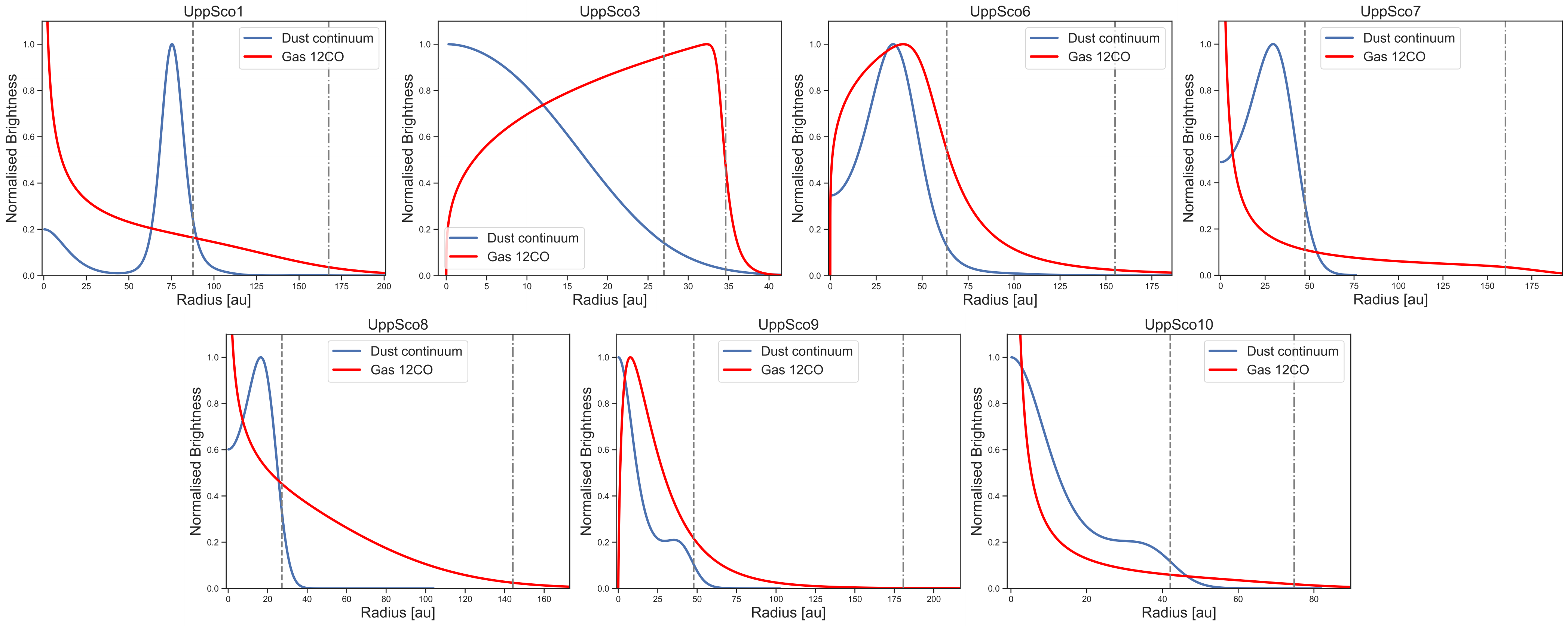}
\caption{Comparison between the Lupus and Upper Scorpius dust-disk radial profiles and gas-disk radial profiles (\textbf{the latter derived} in \citealp{AGEPRO_XI_gas_disk_sizes}). The dust brightness profile is normalized to the peak. Lupus 2, 6, and Upper Scorpius 3, 6, and 9 gas-disks are peak normalized (Sersic profiles). All other gas-disk models (Nuker profiles) go to infinity at the origin, and thus, they are normalized to the same fractional \textbf{arbitrary} intensity. Vertical dashed lines correspond to the dust R\textsubscript{90\%} radii and dotted–dashed lines correspond to R\textsubscript{CO,90\%} radii.}\label{Plot: Lupus_gas_dust}

\end{figure*} 

\subsection{Substructures in Class I Ophiuchus Sources}\label{Oph_substruc}

We have detected a very high fraction of substructures in our Ophiuchus sample ($\sim 80\%$ of the resolved objects). The AGE-PRO Ophiuchus sources were selected by their Class I SED (\citealp{AGEPRO_I_overview}). Analyzing the R\textsubscript{dust}–L\textsubscript{mm} correlation, in Sect. \ref{Sec_population} we have shown that the AGE-PRO Ophiuchus sample is likely placed toward the end of the Class I phase. This agrees with the age estimations presented by the eDisk program (\citealp{2023ApJ...951....8O}). AGE-PRO Ophiuchus 7 (IRS 63) is the oldest source in eDisk (\citealp{2023ApJ...958...98F}), but it is among the youngest sources of AGE-PRO (\citealp{AGEPRO_II_Ophiuchus}), being the only source in common between both surveys. 


It is thus interesting to put the AGE-PRO Ophiuchus sample of Class Is in perspective with other studies of dust substructures in protostellar sources. The eDisk program has studied 12 Class 0 and 7 Class I protostars. They have found that the Band 6 1.3 mm continuum emission barely show any sharp high-contrast features, despite the high angular resolution of the program ($\sim 0.04$"). This is true even in those instances where the visibilities were fitted to try to recover substructures (e.g. \citealp{2024A&A...690A..46S}). I\textbf{n total, only three eDisk sources have substructures reported, including IRS 63 }(\citealp{2020Natur.586..228S}; Sect. \ref{Frank_subsec}, Fig. \ref{Plot: Oph}). A more direct comparison with AGE-PRO is the work of \citet{2023AJ....166..184M}, which employed a visibility modeling analysis similar to the one of Sect. \ref{Frank}, with similar resolution observations. \citet{2023AJ....166..184M} analyzed nine Class Is in Ophiuchus, and proposed substructures for four of them, with another four being ambiguous. Hence, there seems to be an increment in the fraction of substructured disks from Class 0/early Class Is (eDisk), to Class Is (\citealp{2023AJ....166..184M}; \textbf{see also \citealp{2024ApJ...973..138H, 2025arXiv250411577H}}), to late Class Is (AGE-PRO). \citet{2021MNRAS.501.2934C} also observed a much higher incidence of substructures within Ophiuchus toward the sources that have a more Class II SED shape.

One possible explanation for this transition is that substructures appear during the Class I phase, and they are only ubiquitous at the late Class I stage AGE-PRO is tracing. Another explanation is that early Class I objects are too optically thick at 1.3 mm to identify the existing substructures, but quick dust evolution \textbf{or disk-settling} (\citealp{2025A&A...697A..84N}) permits substructures to be more easily detected at the late Class I stage. While substructures might evolve during the Class 0 and Class I stages, the optical thickness explanation is likely majorly responsible for the observed trend. \citet{2024A&A...688A..81D} suggest that substructures appear early (within 0.4 Myr), and \citet{2024A&A...689L...5M} has already found a tentative substructure detection in a Class 0 disk, while substructure detections around young Class I disks are becoming more frequent (e.g. \citealp{2020ApJ...895L...2N,2020ApJ...902..141S,2023AJ....166..184M,2023ApJ...951...11Y,2024ApJ...973..138H,2024ApJ...961..228S}).

We conclude that the very high fraction of substructures detected in the AGE-PRO Ophiuchus sample compared to previous Class I studies is due to its late Class I stage. \textbf{AGE-PRO Ophiuchus} sources had enough time for their disks to become optically thinner, which caused more dust substructures to be recovered. We note that the very high fraction of substructures in the AGE-PRO Ophiuchus sample implies that dust-disk substructures are already ubiquitous at $\sim$ $0.5-1$ Myr. 






\subsection{Radial Profiles: Dust-disk versus Gas-disk}\label{dust-disk vs. gas-disk}



In this section, we compare the dust-disk radial profiles obtained in Sect. \ref{Frank_subsec} with the gas-disk radial profiles derived in \citet{AGEPRO_XI_gas_disk_sizes} from the AGE-PRO \textsuperscript{12}CO moment zero maps \textbf{(using Nuker and Sersic profiles)}. \citet{AGEPRO_XI_gas_disk_sizes} methodology could not be applied to the Ophiuchus sources due to cloud contamination (\citealp{AGEPRO_II_Ophiuchus}). We note that, given the significant differences between both approaches, the comparison between \texttt{Frankenstein} profiles and Nuker profiles presented in this section is only for guidance.

Fig. \ref{Plot: Lupus_gas_dust} shows the dust-disk and gas-disk brightness radial profiles for the well-resolved Lupus and Upper Scorpius sources in the continuum. In all cases, the gas-disk extends farther than the dust-disk (see also \citealp{AGEPRO_XI_gas_disk_sizes}). We see tentative gas cavities on Upper Scorpius  3, 6, and 9 (we note none of them appear resolved in the moment 0 maps, \citealp{AGEPRO_XI_gas_disk_sizes}). Of these, Upper Scorpius 6 (2MASS J16163345-2521505) has a dust cavity with a dust-ring located at a similar distance to where the gas-disk peaks, at around 30-40 au. It is the only\textbf{ AGE-PRO} disk with a large inner dust cavity whose \textsuperscript{12}CO moment zero map is better modeled with a gas cavity. There are well-studied \textit{transition disks} with gas in their dust cavities and those with gas-poor dust cavities (e.g. PDS 70, GM Aur, HD 100546, see \citealp{2023A&A...670A.154W}). \textbf{The presence of gas in the cavity has been associated with multiple Neptunes, with only gas giants being able to carve a gas cavity (\citealp{2024A&A...691A.155H}).} We hence pose Upper Scorpius 6 as a disk with a gas-cleared inner dust cavity. Upper Scorpius 3 is one of the most compact AGE-PRO gas-disks, so the gas cavity might be an artifact from \citet{AGEPRO_XI_gas_disk_sizes} methodology. We note that for Upper Scorpius 3 the dust-disk radii and gas-disk radii are similar. As for Lupus 9 (Appendix \ref{final word}), a companion truncating the gas-disk is a possible explanation. Upper Scorpius 9 shows a small gas cavity extending to 10 au. This might suggest Upper Scorpius 9 has an unresolved dust cavity at the 10 au scale (supporting the discussion of Sects. \ref{Sec_population} and \ref{discuss_structures}). 





The other {disks with inner dust cavities} Upper Scorpius 1, Upper Scorpius 7, and Upper Scorpius 8 show very extended gas-disks, extending to similar sizes $\sim 145-165$ au, independently of the position of the dust-ring (e.g., at 78 au for Upper Scorpius 1 vs. at 17 au for Upper Scorpius 8). More information on the relation between gas and dust structure in Upper Scorpius 1 can be found in \citet{2024ApJ...974..102S}, where hints of a protoplanet in the dust cavity are presented. The colossal extent of Lupus 10 (V1094 Sco) is worth mentioning, extending to $\sim833$ au in \textsuperscript{12}CO gas. 



\section{Conclusions}\label{Conclusions2}



We performed visibility fitting to the Band 6 1.3 mm continuum data of the 30 AGE-PRO  ALMA Large Program protoplanetary disks (\citealp{AGEPRO_I_overview}). AGE-PRO contains 10 disks in Ophiuchus \textbf{($0.5$-$1$ Myr)}, 10 disks in Lupus \textbf{($1$-$3$ Myr)}, and 10 disks in Upper Scorpius \textbf{($2$-$6$ Myr)}, all in a narrow stellar mass range (0.3-0.8 M$_{\odot}$). \textbf{ We derived} disk geometries (inclinations and position angles, Table \ref{Table1}) \textbf{for the AGE-PRO protoplanetary disks}. For all sources, we homogeneously obtained dust-disk radii (R\textsubscript{68\%}, R\textsubscript{90\%}\textbf{, and R\textsubscript{95\%},} Table \ref{Table2}), and azimuthally symmetric radial profiles of the intensity of the dust continuum emission (Figs. \ref{Plot: Oph}, \ref{Plot: Lupus}, \ref{Plot: UppSco}).  We showed that the AGE-PRO sample is representative of the considered regions (Sect. \ref{analysis}). The mean resolution elements achieved in the visibility fitting are $\sim$13, 22, and 22 au for Ophiuchus, Lupus, and Upper Scorpius sources, respectively (Sect. \ref{Frank}). 

We examined the presence of continuum substructures by using the radial profiles and their residuals (Fig. \ref{Interest_residuals_mosaic}). Then, we analyzed the evolution of dust-disk radii and substructures across Ophiuchus, Lupus, and Upper Scorpius. The main conclusions we have obtained are as follows:


\begin{itemize}
    
    \item We identify substructures in 50\% of the AGE-PRO disks (seven in Ophiuchus, two in Lupus, and six in Upper Scorpius). Our methodology detects substructure in almost all sources \textbf{($\sim$$90\%$)} observed at high enough resolution for their disk size ($\theta_{D}/\theta_{res}>3$). We report five disks with large ($>$15 au) inner dust cavities and a ring (or \textit{transition disks}), four of them in the Upper Scorpius sample. This suggests a trend toward disk populations with higher fractions of disks with large inner dust cavities at later ages.

    \item We detect no R\textsubscript{dust}–L\textsubscript{mm} significant correlation for the AGE-PRO Upper Scorpius sample. This can be reconciled with previous \textbf{literature} results considering the narrow mass range and the small size of the AGE-PRO sample. For Ophiuchus and Lupus, we recover L\textsubscript{mm}$\sim R^{2}$ (Fig. \ref{Plot: FLux_Radii}). We see no evolution of dust-disk radii with time in the AGE-PRO sample (Fig. \ref{Plot: radii_compa_2}). We suggest that a high fraction of disks in Upper Scorpius with dust trapping substructures is the best explanation for these two results.

    \item{Upper Scorpius disks with large inner dust cavities (or \textit{transition disks}) have gas-disk masses compatible with the median gas-disk mass of the Lupus and Upper Scorpius regions (Fig. \ref{Plot: Mgas_vs_Rdust}). In contrast, Lupus and Upper Scorpius substructured disks with no large dust cavities\textbf{ have gas-disk masses 1 to 2 orders of magnitude higher} than the average values in their regions. This suggests that at the age of $\sim$$1$-$2$ Myr a significant fraction of disks have developed dust traps which are similarly effective to large inner dust cavities to remove gas mass. \textbf{A different interpretation is to explain the large gas-disk masses} via late infall events and interaction with the environment.}


    \item{A spiral is identified in IRS 63 (Fig. \ref{Interest_residuals_mosaic}), hinting to gravitational instability in this massive disk.}

    \item{The AGE-PRO Ophiuchus Class I disks show dust-disk substructures in $\sim$80\% of the well-resolved sources (Fig. \ref{Plot: Oph}). This evidences the early formation of substructures in protoplanetary disks, which appear to be ubiquitous at $0.5$-$1$ Myr. Using the R\textsubscript{dust}–L\textsubscript{mm} relationship, we deduce the AGE-PRO Ophiuchus sample is at the end of the Class I phase, reconciling our results with other Class I studies that detect a lower incidence of substructures at earlier ages, and suggesting an evolutionary trend or optical thickness effects (Sect. \ref{Oph_substruc}).} 

    \item{We present a comparison of the radial intensity profiles of the dust continuum emission reported in this work with the gas-disk radial profiles derived in \citet{AGEPRO_XI_gas_disk_sizes} from the AGE-PRO \textsuperscript{12}CO moment zero maps. We discuss colocal substructures in both tracers.}
    
\end{itemize}

\textbf{Considering the above results,} we propose an evolutionary sequence in which protoplanets at inner orbits have time to open dust cavities in $\sim$$1$-$2$ Myr. This \textbf{results in} disk populations \textbf{with} a higher fraction of large inner dust cavities at later ages. In turn, this fixes the outer dust-disk radii in a fraction of the population for which the dust-radius does not evolve over time and causes the absence of significant R\textsubscript{dust}-L\textsubscript{mm} correlation in the AGE-PRO Upper Scorpius sample. This evolutionary scenario also explains why disks with large inner dust cavities have gas-masses typical of their region, the dust modeling results of \citet{AGEPRO_VI_DustEvolution}, and the absence of evolution of the $R_{\rm CO,\ 90\%}/R_{\rm dust,\ 90\%}$ ratio \textbf{in the AGE-PRO sample} from Lupus to Upper Scorpius (\citealp{AGEPRO_XI_gas_disk_sizes}). This evolutionary scenario would imply a large fraction of  protoplanetary disks with inner planets \textbf{capable of creating large dust cavities}, of at least 40\% \textbf{of the disk-hosting Upper Scorpius population}, based on our limited sample size statistics. 

This AGE-PRO work constitutes the first analysis of the evolution of dust-disk radii and dust substructures with a representative sample of protoplanetary disks in three different star forming regions. A detailed discussion for each individual object is presented in Appendix \ref{final word}.



\section{Acknowledgements}

We thank Álvaro Ribas, Osmar Guerra-Alvarado, and John Tobin for insightful discussions that improved this work. A.S. acknowledges support from FONDECYT de Postdoctorado 2022 \#3220495 and support from the UK Research and Innovation (UKRI) under the UK government’s Horizon Europe funding guarantee from ERC (under grant agreement No. 101076489). G.R. acknowledges funding from the Fondazione Cariplo, grant No. 2022-1217, and the European Research Council (ERC) under the European Union’s Horizon Europe Research \& Innovation Programme under grant agreement No. 101039651 (DiscEvol). Views and opinions expressed are however those of the author(s) only, and do not necessarily reflect those of the European Union or the European Research Council Executive Agency. Neither the European Union nor the granting authority can be held responsible for them. K.Z. acknowledges the support of the NSF AAG grant No. 2205617. L.P. acknowledges support from ANID BASAL project FB210003 and ANID FONDECYT Regular \#1221442. L.T. acknowledges the support of NSF AAG grant No. 2205617. L.A.C. acknowledges support from the Millennium Nucleus on Young Exoplanets and their Moons (YEMS), ANID - Center Code NCN2021\_080 and the FONDECYT grant No. 1241056. C.A.G. acknowledges support from FONDECYT de Postdoctorado 2021 \#3210520. P.P acknowledges the support from the UK Research and Innovation (UKRI) under the UK government’s Horizon Europe funding guarantee from ERC (under grant agreement No. 101076489). B.T. acknowledges the support of the Programme National Physique et Chimie du Milieu Interstellaire (PCMI) of CNRS/INSU with INC/INP cofunded by CEA and CNES. C.G-R. acknowledges support from the Millennium Nucleus on Young Exoplanets and their Moons (YEMS), ANID - Center Code NCN2021\_080. I.P. and D.D. acknowledge support from Collaborative NSF Astronomy \& Astrophysics Research grant (ID: 2205870). P.C. acknowledges support by the Italian Ministero dell’Istruzione, Universit\`a e Ricerca through the grant Progetti Premiali 2012 – iALMA (CUP C52I13000140001) and by the ANID BASAL project FB210003. J.M acknowledges support from FONDECYT de Postdoctorado 2024 \#3240612, J.M. acknowledges support from the Millennium Nucleus on Young Exoplanets and their Moons (YEMS), ANID - Center Code NCN2021\_080. N.T.K. and P.P. acknowledge the support provided by the Alexander von Humboldt Foundation in the framework of the Sofja Kovalevskaja Award endowed by the Federal Ministry of Education and Research.

This paper makes use of the following ALMA data: 
ADS/JAO.ALMA\#2016.1.00484.L, \\
ADS/JAO.ALMA\#2017.1.01167.S, \\
ADS/JAO.ALMA\#2018.1.00028.S, \\
ADS/JAO.ALMA\#2018.1.00271.S, \\
ADS/JAO.ALMA\#2019.1.00738.S, \\
ADS/JAO.ALMA\#2021.1.00128.L. \\
ALMA is a partnership of ESO (representing its member states), NSF (USA) and NINS (Japan), together with NRC (Canada), MOST and ASIAA (Taiwan), and KASI (Republic of Korea), in cooperation with the Republic of Chile. The Joint ALMA Observatory is operated by ESO, AUI/NRAO, and NAOJ. The National Radio Astronomy Observatory is a facility of the National Science Foundation operated under cooperative agreement by Associated Universities, Inc.

\newpage
\appendix

\section{Disk Geometries from the Gas-disk}\label{Leon_work_appendix}

Using the method described in \citet{AGEPRO_XI_gas_disk_sizes}, we fit two-dimensional Gaussians to the \textsuperscript{12}CO moment zero maps of the Lupus and Upper Scorpius sources. This provides a fit to the gas-disk geometry. This methodology could not be applied to the Ophiuchus sources due to the cloud contamination their \textsuperscript{12}CO maps show (\citealp{AGEPRO_II_Ophiuchus}). For well-resolved sources in the continuum visibilities (Sect. \ref{galario}), there is {agreement within error bars} between the disk geometries obtained from this gas analysis and the dust geometries obtained in Sect. \ref{galario} with \texttt{Galario} plus MCMC fit. Hence, for all well-resolved sources, we opt for the dust-based geometries as their uncertainties are smaller, and they are obtained from a direct fit to the visibilities. An exception to this is Upper Scorpius 3. For Upper Scorpius 3, the \texttt{Galario}-based disk geometry results in large residuals for their best \texttt{Frankenstein} model. The fit to the \textsuperscript{12}CO-based geometry does not show these residuals, and we therefore adopt the \textsuperscript{12}CO-based geometry for Upper Scorpius 3 \textbf{in this work} (Table \ref{tab: gasfit disk geometry}). 



\begin{figure}[ht!]
     \centering
     \includegraphics[width=0.5\columnwidth]{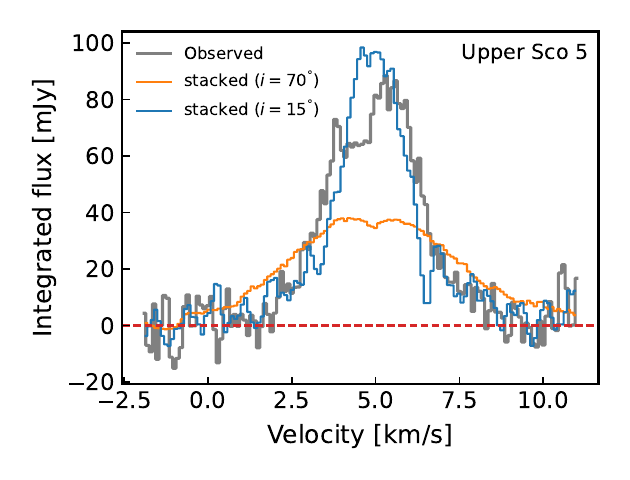}
     \caption{\label{fig: UpperSco 5 spectrum} \textsuperscript{12}CO 2–1 integrated spectrum for Upper Scorpius 5 (2MASS J16145026-2332397). The gray line shows the observed spectrum (see \citealp{AGEPRO_IV_UpperSco}). Shown in orange is a velocity-stacked spectrum (made using \texttt{GoFish}, \citealp{2019JOSS....4.1632T}) using the position angle and inclination from the continuum disk geometry fit (Sect. \ref{galario}). The blue line shows a velocity-stacked spectrum assuming $\text{PA}=139^{\circ}$ and $\text{inc}=15^{\circ}$, the disk geometry assumed in this work.} 
\end{figure}

For the eight marginally resolved sources in continuum visibility space (Sect. \ref{galario}), we tried to adopt the achieved geometric fit to the \textsuperscript{12}CO data. Gas emission is often more extended than dust \textbf{emission}, thereby providing an additional and often more reliable constraint on the disk geometry in these cases where the dust-disks are only marginally resolved (e.g. Fig. \ref{fig: UpperSco 5 spectrum}). However, in Ophiuchus 5 and Lupus 8, cloud contamination impedes us from reaching a good \textsuperscript{12}CO fit. Lupus 9 and Upper Scorpius 2 appear marginally resolved in \textsuperscript{12}CO gas.  For Upper Scorpius 4 we obtain similar results between the \textsuperscript{12}CO-fit and the continuum-based fit. Hence, for these five latter sources, we opt to use the dust-based geometry.

\begin{table}[ht!]
\setlength{\tabcolsep}{2pt}
\caption{\label{tab: gasfit disk geometry} Disk Geometries Obtained in Sect. \ref{galario} from Modeling the Continuum Visibilities and Disk Geometries Obtained in Sect. \ref{Leon_work} from Fitting the \textsuperscript{12}CO Moment Zero Maps}
\def\arraystretch{1.3}
\begin{tabular*}{0.5\columnwidth}{l|cc|cc}
\hline\hline
      & \multicolumn{2}{c}{From \textsuperscript{12}CO} & \multicolumn{2}{c}{From Continuum}\\
Name  &  PA   & Inc    & PA    & Inc\\
      & (deg) & (deg)  & (deg) & (deg)\\
\hline
Lupus 4   & $20.0 \pm 0.7$ & $31.5 \pm 0.6$ & $46.2^{+8.5}_{-9.7}$ &     $58.4^{+9.7}_{-8.3}$ \\
Lupus 5   & $8.2 \pm 2.1$ & $26.0 \pm 0.8$ & $108.0^{+34}_{-34}$ &     $19.8^{+13}_{-13}$ \\
Upper Scorpius 3  &  $78.1\pm 9.6$ & $58.2\pm0.1$ & $79.3^{+1.4}_{-1.4}$ &     $72.0^{+2.1}_{-2.2}$ \\
Upper Scorpius 5$^{\dagger}$  & $139\pm 5.3$ & $15\pm10$ &  $68.6^{+9.8}_{-10.3}$ &   $70.6^{+13.2}_{-16.3}$ \\
\hline\hline
\end{tabular*}
\begin{minipage}{0.5\columnwidth}
\vspace{0.18cm}
{\footnotesize{{Notes:} We only show the sources for which the \textsuperscript{12}CO geometries were considered for subsequent analysis in this work. $^{\dagger}$: The disk geometry of Upper Scorpius 5 is obtained from minimizing the width of the velocity-stacked \textsuperscript{12}CO spectrum obtained with \texttt{GoFish} (\citealp[see Figure \ref{fig: UpperSco 5 spectrum}]{2019JOSS....4.1632T}).}}
\vspace{0.1cm}
\end{minipage}
\end{table}

For 2MASS J16145026-2332397 (Upper Scorpius 5) we took a different approach. An inspection of the \textsuperscript{12}CO line profile shows a nearly face-on disk (see Fig. \ref{fig: UpperSco 5 spectrum} and \citealp{AGEPRO_IV_UpperSco}). Velocity-stacking the spectrum according to the continuum-fitted disk geometry results in a very broad spectrum rather than the narrow spectrum we expect from a correctly velocity-stacked Keplerian disk. Experimenting with different disk geometries, we find that (PA, inc) = ($139^{\circ}$, $15^{\circ}$) yields a narrow velocity-stacked spectrum. We therefore adopt these numbers \textbf{in this work} for the disk geometry of Upper Scorpius 5 (Table \ref{tab: gasfit disk geometry}).


In conclusion, only for four sources (Lupus 4, 5, Upper Scorpius 3, and 5) the \textsuperscript{12}CO-based geometries are used as a starting point for the \texttt{Frankenstein} fits of Sect. \ref{Frank} and subsequent estimation of the dust-disk radii. These cases where geometries are derived from gas fits rather than from continuum visibilities are summarized in Table \ref{tab: gasfit disk geometry}. The final adopted geometric properties of the entire sample are given in Table \ref{Table1}.

\section{Comparison with Previous Literature}\label{pre_lit}

Among the AGE-PRO sample, there are seven sources with publicly available observations \textbf{with higher angular resolution than AGE-PRO}. \textbf{These observations} can be compared with the visibility models of the dust continuum emission \textbf{derived in this work}. The Lupus 2 system was observed as part of DSHARP \citep{Andrews_DSHARP_2018ApJ...869L..41A}; Ophiuchus 1 and Ophiuchus 7 are in the ODISEA high-resolution sample \citep{2021MNRAS.501.2934C}; Lupus 1 and Lupus 8 are a binary system and were published in \citet{2024A&A...682A..55M}, while Lupus 3 and Upper Scorpius 10 have not been published before. \textbf{All sources have Band 6 1.3 mm data, except for Lupus 3, which has Band 7 0.9 mm data.}

\begin{figure*}
\includegraphics[width=\textwidth]{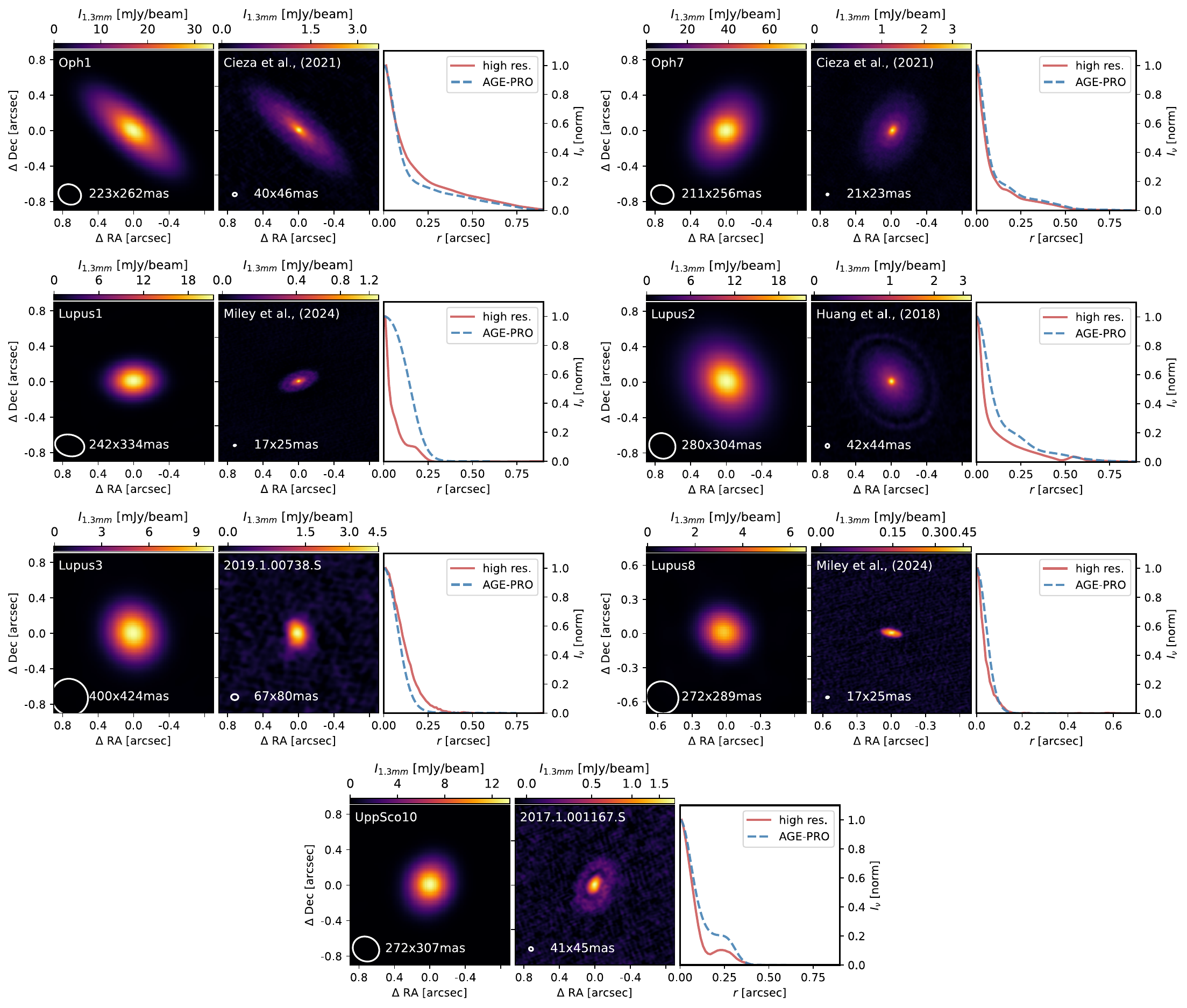}
    \caption{Comparison of the radial profiles \textbf{obtained in} this work from visibility fitting (Sect. \ref{Frank_subsec}), with the radial profiles from publicly available high angular resolution images at comparable wavelengths (first two panels show the CLEAN images of each dataset).}
    \label{Plot: HR_comp}
\end{figure*}

The detection of dust substructures is highly dependent on resolution and sensitivity. Hence, it is challenging to compare substructures from different observations. For the sources that have been already published, we compare our visibility-modeled profiles to the self-calibrated dust continuum images at high angular resolution. \textbf{To obtain radial profiles from the images, we take azimuthal averages.} For the sources that have not been published before, we compare\textbf{ our visibility-modeled profiles to} the images that are available from the ALMA archive (thus not self-calibrated). We show these comparisons in Fig.~\ref{Plot: HR_comp}. We note that fits to the visibilities cannot be compared with CLEAN images in a fair manner, and that the comparisons presented in this appendix are only for guidance.

The visibility modeled \textbf{intensity radial} profiles of this work are in very good agreement with the disk’s intensity shapes from the high-resolution images. Ophiuchus 1, 7, Lupus 1, Lupus 2, and Upper Scorpius 10 have low-contrast dust continuum substructures in the high-resolution data. Even though those substructures are not detected in the CLEAN images of AGE-PRO, the visibility modeled profiles of this work do retrieve the presence of those substructures at the right locations (Fig.~\ref{Plot: HR_comp}). This is true for all cases but Lupus 1, which is in our `not-structured' sample (Fig. \ref{Plot: Lupus}, Table \ref{Table2}). Lupus 3 and Lupus 8 are in our `not-structured' sample and `marginally resolved' sample, respectively, and they show a smooth profile even at the higher resolution. In Upper Scorpius 10, a shoulder-like emission in the visibility model resolves into a faint ring when observed at higher angular resolution. For Ophiuchus 1, the \texttt{Frankenstein} substructure retrieved in this work appears more evident than in the high-resolution image. \textbf{This is likely due to the high inclination of Ophiuchus 1, which causes that even at higher angular resolutions the beam smears significantly the substructures in the image.}






This comparison with higher-resolution data evidences the power of visibility fitting to
identify dust substructures in medium-resolution data (typical CLEAN beam size of AGE-PRO is $\sim0.2$-$0.35$", in comparison to the typical \texttt{Frankenstein} beam of $\sim0.09$-$0.15$"). In addition, this comparison shows that we can trust the brightness radial profiles obtained in Sect. \ref{Frank_subsec} from visibility modeling for those sources in which we do not have high angular resolution available.


\section{Discussion on Individual Sources}\label{final word}

\begin{figure*}[t!]
\includegraphics[width=\textwidth]{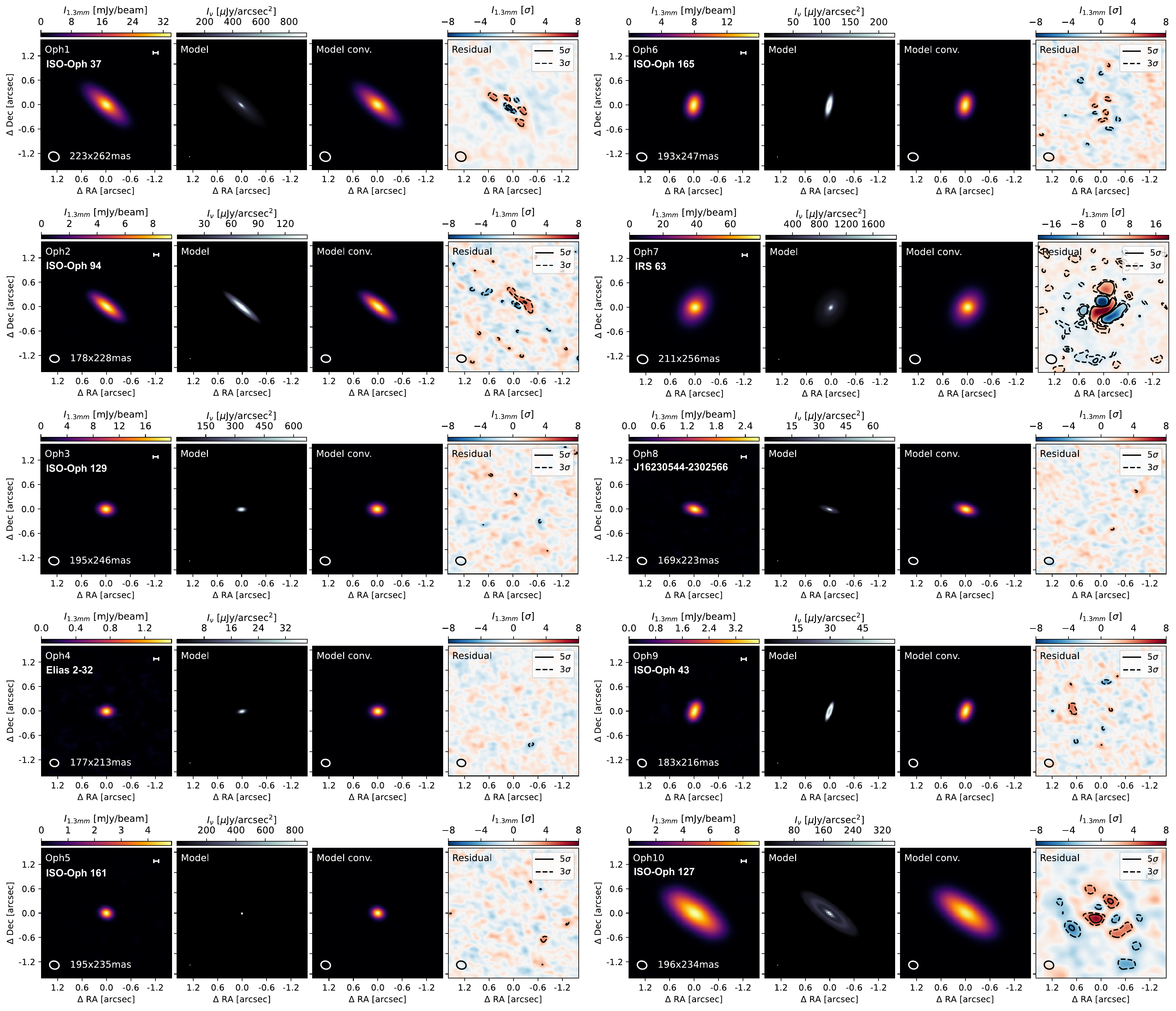}
\caption{Mosaic \textbf{of the residuals of AGE-PRO} Ophiuchus sources. For each source, we show the CLEAN image of the continuum-only data, the image of the model with pixel size, the image of the model convolved with the CLEAN beam, and the residual image from subtracting the visibilities of the data from those of the model. Residuals are shown in units of the rms noise value.}\label{Oph_residuals_mosaic}
\end{figure*} 

\begin{figure*}[t!]
\includegraphics[width=\textwidth]{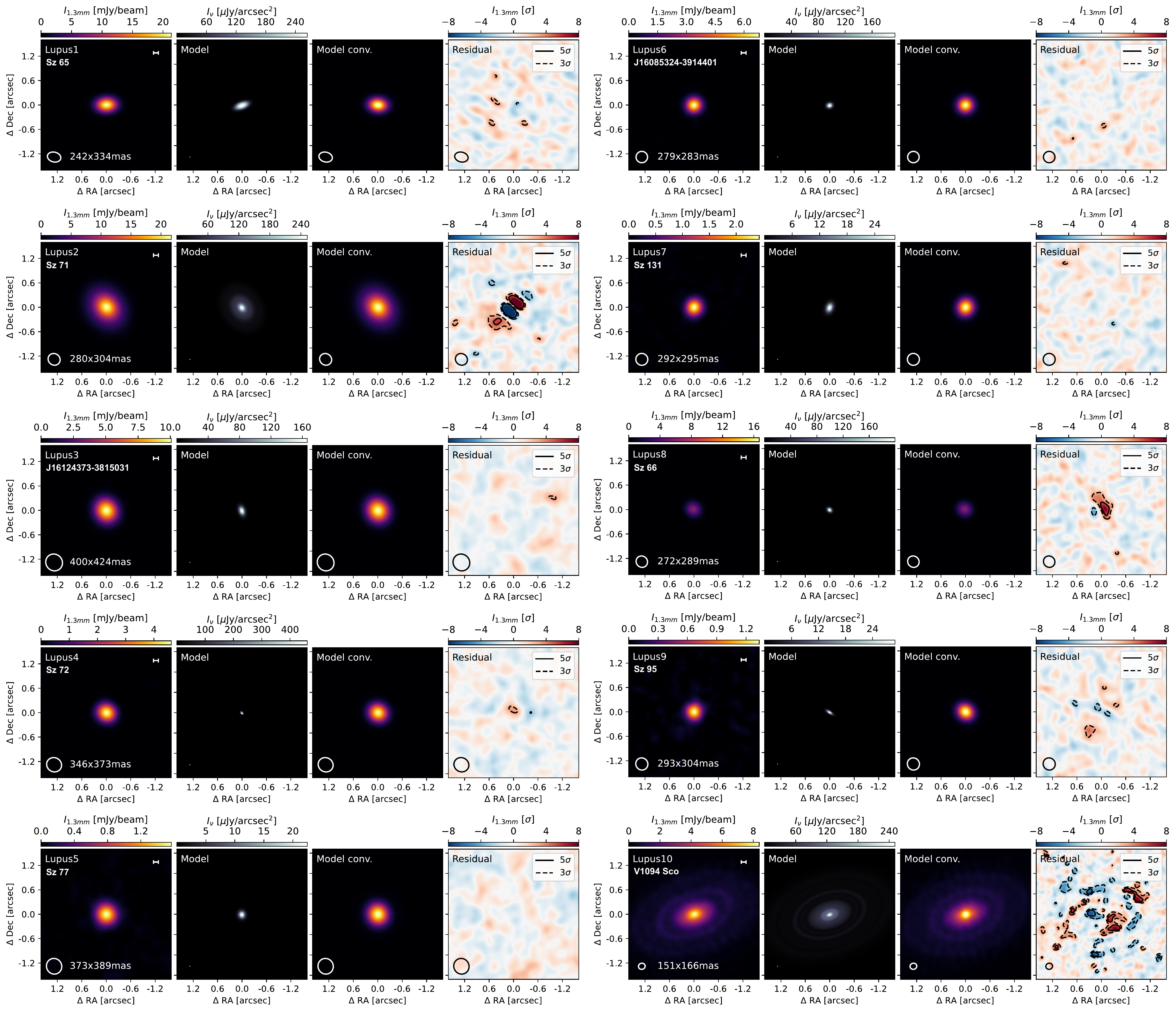}
\caption{Mosaic \textbf{of the residuals of AGE-PRO} Lupus sources. For each source, we show the CLEAN image of the continuum-only data, the image of the model with pixel size, the image of the model convolved with the CLEAN beam, and the residual image from subtracting the visibilities of the data from those of the model. Residuals are shown in units of the rms noise value.}\label{Lupus_residuals_mosaic}
\end{figure*} 

\begin{figure*}[t!]
\includegraphics[width=\textwidth]{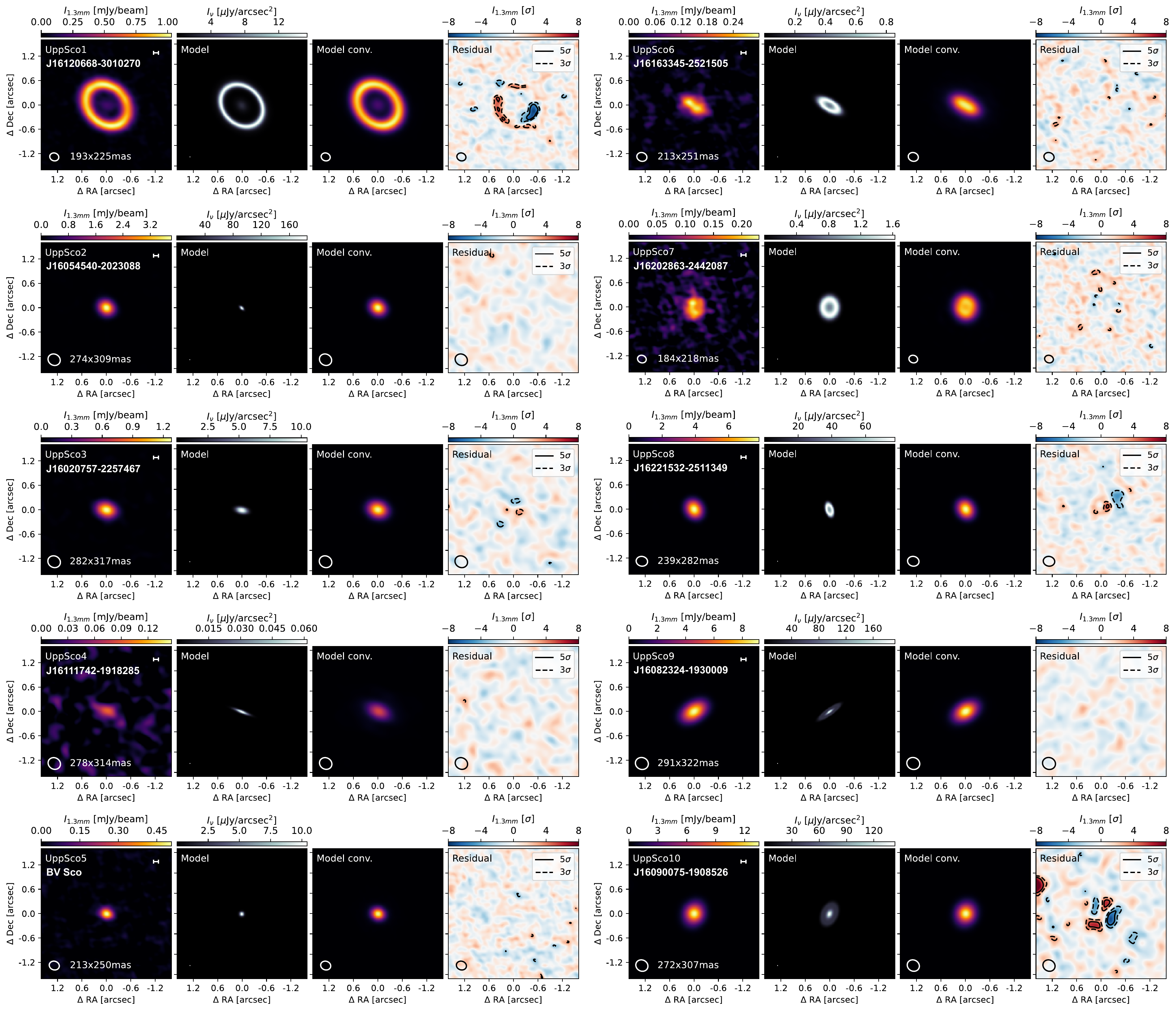}
\caption{Mosaic \textbf{of the residuals of AGE-PRO} Upper Scorpius sources. For each source, we show the CLEAN image of the continuum-only data, the image of the model with pixel size, the image of the model convolved with the CLEAN beam, and the residual image from subtracting the visibilities of the data from those of the model. Residuals are shown in units of the rms noise value.}\label{UppSco_residuals_mosaic}
\end{figure*} 


In this appendix, we summarize the final results of this work for each AGE-PRO protoplanetary disk. Individual radial profiles of the dust continuum emission are presented in Figs. \ref{Plot: Oph}, \ref{Plot: Lupus}, and \ref{Plot: UppSco} for Ophiuchus, Lupus, and Upper Scorpius sources, respectively. The obtained residuals are shown in Figs. \ref{Oph_residuals_mosaic}, \ref{Lupus_residuals_mosaic}, and \ref{UppSco_residuals_mosaic}, for Ophiuchus, Lupus, and Upper Scorpius sources, respectively. Disk geometries and dust-disk radii for every source are compiled in Tables \ref{Table1} and \ref{Table2} with their uncertainties. Detailed information on gas-disk radii and gas residuals can be found in \citet{AGEPRO_XI_gas_disk_sizes}. 



\subsection{Ophiuchus Sources}

\textit{SSTc2d J162623.6-242439 (Ophiuchus 1, ISO-Oph 37, $\text{inc}=72.59^{\circ}$, $R_{90\%}=95$ au).} The dust-disk is large, extending to a R\textsubscript{90\%} of $95$ au. Residuals show one 5$\sigma$ small central area and several 3$\sigma$ spots unevenly distributed across the dust-disk. \citet{2021MNRAS.501.2934C}, using higher-resolution data, finds no substructures, except for an inflection point at 31 au also identified in this work (at $\sim$20 au, Fig. \ref{Plot: Oph}). Due to its lower spatial resolution, the AGE-PRO brightness temperature sensitivity is higher than in the previous high-resolution observations. Hence, we propose that these residuals might be signaling the existence of a real substructure previously hidden by strong optical thickness in this young Class I source. The symmetric shape of the residuals could be indicative of an underlying dust trap or be related to a snowline.

\textit{SSTc2d J162703.6-242005 (Ophiuchus 2, ISO-Oph 94, $\text{inc}=76.40^{\circ}$, $R_{90\%}=69$ au).} This is a structured disk with a flat top brightness distribution of the radial profile until 20 au and a \textbf{shoulder} at 60 au. Residuals show a 3$\sigma$ asymmetric structure concentrated to the \textbf{northeast} of the disk. Small localized 5$\sigma$ emission is found at the disk center and in the residuals to the \textbf{northeast}. These \textbf{residuals seem to trace real structure} (e.g. \citealp{2021ApJ...916...51A}), and we propose vertical height in the dust distribution (e.g. \citealp{2024MNRAS.532.1752R}) or a flared disk as possible physical explanations.

\textit{SSTc2d J162719.2-242844 (Ophiuchus 3, ISO-Oph 129, $\text{inc}=61.25^{\circ}$, $R_{90\%}=17.5$ au).} There is no substructure at the resolution of AGE-PRO. The residuals show no significant extra emission. 

\textit{SSTc2d J162728.4-242721 (Ophiuchus 4, Elias 2-32, $\text{inc}=61.0^{\circ}$, $R_{90\%}=18.1$ au).} There is no obvious substructure at the resolution of AGE-PRO. The residuals show no significant extra emission. The visibilities show deviations that no \texttt{Frankenstein} model fitted, which might be indicative of underlying substructures.

\textit{SSTc2d J162737.2-244237 (Ophiuchus 5, ISO-Oph 161, $\text{inc}=25^{\circ}$,  $R_{90\%}$ upper limit of $5.5$ au).} This disk is marginally resolved in the visibility fitting. The residuals show no significant extra emission. 

\textit{SSTc2d J162738.9-244020 (Ophiuchus 6, ISO-Oph 165, $\text{inc}=74.34^{\circ}$, $R_{90\%}=40.1$ au).} This is a disk with a flat top brightness distribution of the radial profile until 20 au followed by a smooth decay. Residuals show evenly distributed 3$\sigma$ emission at the disk location.

\textit{SSTc2d J163135.6-240129 (Ophiuchus 7, IRS 63, $\text{inc}=46.94^{\circ}$, $R_{90\%}=74$ au).} This is a structured disk with inflection points in the dust-continuum radial profile at 10, 30, and 80 au. Residuals show a newly identified spiral, adding further evidence to the suggestion that this disk is gravitationally unstable. \citet{2023AJ....166..184M} already found residuals highlighting nonaxisymmetric excess emission just south and north of the disk center. This source has been the object of detailed studies (\citealp{2020Natur.586..228S,2021MNRAS.501.2934C,2023ApJ...958...98F,2023AJ....166..184M,2024A&A...688L..22P}). We believe this spiral was not detected before because it is a low-contrast structure that can be hidden at high resolution. The contrast between the peak intensity in the CLEAN image and the peak intensity in the spiral is of the order of 1\%, so deep observations are needed to trace changes in density, where the spiral lives. Only the deep AGE-PRO observations have allowed us to see the spiral, indicating it must be very optically thick at 1.3 mm. Follow-up analyses are needed to better characterize this spiral. 

\textit{2MASS J16230544-2302566 (Ophiuchus 8, $\text{inc}=71.84^{\circ}$, $R_{90\%}=34.5$ au).} This is a substructured disk with a shoulder in the intensity radial profile at 20-30 au. The residuals show no significant extra emission. 

\textit{SSTc2d J162627.5-244153 (Ophiuchus 9, ISO-Oph 43, $\text{inc}=73.48^{\circ}$, $R_{90\%}=32.3$ au).} This is a newly identified \textbf{\textit{transition disk} with a dust cavity} and dust-ring at 20-25 au. It appears as a transition disk candidate in \citet{2016A&A...592A.126V}, although it was not identified as a transition disk in \citet{2017ApJ...851...83C} or \citet{2019MNRAS.482..698C}. The residuals show no significant extra emission on the disk location, but there is a 3$\sigma$ positive residual to the west of the source. 

\textit{SSTc2d J162718.4-243915 (Ophiuchus 10, ISO-Oph 127, $\text{inc}=73.62^{\circ}$, $R_{90\%}=54.3$ au).} The intensity of the radial profile shows a structured decay with a prominent ring at 30 au followed by a secondary ring at 55 au. We find 3$\sigma$ residuals covering the extent of the disk with localized 5$\sigma$ residuals to the east and south. These residuals are likely not real, and we believe they are produced by an offset in the brightness of the model.

\subsection{Lupus Sources}

\textit{Sz 65 (Lupus 1, IK Lup, $\text{inc}=61.41^{\circ}$, $R_{90\%}=34.10$ au, $R_{CO, 90\%}= 162$ au).} There is no substructure at the resolution of AGE-PRO. The residuals show no significant extra emission. However, we note it appears as slightly structured in the high-resolution work of \citet{2024A&A...682A..55M}. It is the binary of Lupus 8.

\textit{Sz 71 (Lupus 2, GW Lup, $\text{inc}=38.36^{\circ}$, $R_{90\%}=91.1$ au, $R_{CO, 90\%}= 310$ au).} This is a slightly structured disk with a small bump in the intensity radial profile at 40 au. The dust-disk is large, extending to a R\textsubscript{90\%} of $91.1$ au. This system was observed as part of DSHARP ALMA Large Program \citep{Andrews_DSHARP_2018ApJ...869L..41A}, with a substructure matching ours. We find strong bipolar residuals. Because it is a bright disk, we believe that a small phase center deviation is the most likely explanation for these bipolar residuals (\citealp{2021ApJ...916...51A}). Two-dimensional Gaussians to the \textsuperscript{12}CO moment zero maps report a similar inclination to the one used in this work. This inclination is also similar to the one found in DSHARP (\citealp{2018ApJ...869L..42H}).

\textit{2MASS J16124373-3815031 (Lupus 3, $\text{inc}=51.83^{\circ}$, $R_{90\%}=31.5$ au, $R_{CO, 90\%}= 108$ au).}  There is no substructure at the resolution of AGE-PRO. This agrees with archival unpublished data at higher resolution. The residuals show no significant extra emission. 

\textit{Sz 72  (Lupus 4, HM Lup, $\text{inc}=31.50^{\circ}$, $R_{90\%}$ upper limit of $12$ au, $R_{CO, 90\%}= 30$ au).} This disk is marginally resolved in the visibility plane. It shows a faint not centered 3$\sigma$ residual.

\textit{Sz 77 (Lupus 5, $\text{inc}=26.0^{\circ}$, $R_{90\%}$ upper limit of $28$ au, $R_{CO, 90\%}= 51$ au).} This disk is marginally resolved in the visibility plane. The residuals show no significant extra emission. 

\textit{2MASS J16085324-3914401 (Lupus 6, $\text{inc}=32.2^{\circ}$, $R_{90\%}=21.8$ au, $R_{CO, 90\%}=53$ au).} There is no substructure at the resolution of AGE-PRO. The residuals show no significant extra emission. The visibilities show deviations that no \texttt{Frankenstein} model fitted, which might be indicative of underlying substructures. \citet{AGEPRO_XI_gas_disk_sizes} proposes it might have a close companion from the \textsuperscript{12}CO moment zero maps.

\textit{Sz 131 (Lupus 7, $\text{inc}=46.7^{\circ}$, $R_{90\%}=30.4$ au, $R_{CO, 90\%}=72$ au).} There is no substructure at the resolution of AGE-PRO. The residuals show no significant extra emission. 

\textit{Sz 66 (Lupus 8, $\text{inc}=76.8^{\circ}$, $R_{90\%}$ upper limit of $21$ au, $R_{CO, 90\%}=74$ au).} This disk is marginally resolved in the visibility plane. It appears as not structured in the high-resolution work of \citet{2024A&A...682A..55M}, only showing slight evidence of an asymmetric dust distribution when the visibilities were modeled. We believe the strong residuals we detect for this source are entirely caused by the AGE-PRO calibration pipeline on a very faint object. It is the binary of Lupus 1. 

\textit{Sz 95 (Lupus 9, $\text{inc}=54.64^{\circ}$, $R_{90\%}$ upper limit of $31$ au, $R_{CO, 90\%}=6.4$ au).}  This disk is marginally resolved in the visibility plane. The residuals show no significant extra emission in the disk. However, the residuals show a localized 3-5$\sigma$ spot to the southwest of the disk, outside of the gas-disk ($R_{CO, 90\%}=0.04$" or $6.4$ au). This source is among the smallest dust-disk and gas-disk sources of AGE-PRO. This 3-5$\sigma$ continuum residual outside the edge of the gas-disk might be tracing a binary that is truncating the disk.

\textit{V1094 Sco (Lupus 10, $\text{inc}=52.02^{\circ}$, $R_{90\%}=353.0$ au, $R_{CO, 90\%}= 833$ au).} This is a very large, CO-bright (\citealp{2023ApJ...953..183P}), and structured dust-disk, extending to a R\textsubscript{90\%} of $353.0$ au. Several significant oscillations \textbf{(rings and gaps)} can be seen in the intensity radial profile. Residuals show positive and negative 5$\sigma$ detections at separated locations, which overlap with positions of dust-rings (Fig. \ref{Plot: Lupus} and \citealp{2018A&A...616A..88V}). The western positive residual might be tracing a spiral arm.

\subsection{Upper Scorpius Sources}

\textit{2MASS J16120668-3010270 (Upper Scorpius 1, $\text{inc}=37.00^{\circ}$, $R_{90\%}=87.64$ au, $R_{CO, 90\%}= 168$ au).} This is a newly identified \textit{transition disk}, with a large inner dust-cavity and a dust-ring at 78 au. Residuals show a spiral arm structure discussed in detail in \citet{2024ApJ...974..102S}, which also proposes a tentative protoplanet detection within the dust cavity.

\textit{2MASS J16054540-2023088 (Upper Scorpius 2, $\text{inc}=56.4^{\circ}$, $R_{90\%}$ upper limit of $17$ au, $R_{CO, 90\%}= 51$ au).} This disk is marginally resolved in the visibility plane. The residuals show no significant extra emission. 

\textit{2MASS J16020757-2257467 (Upper Scorpius 3, $\text{inc}=58.20^{\circ}$, $R_{90\%}=27.0$ au, $R_{CO, 90\%}= 34$ au).} There is no substructure at the resolution of AGE-PRO. 3$\sigma$ residuals appear at the disk location. It shows a \textbf{tentative} inner gas cavity (Fig. \ref{Plot: Lupus_gas_dust}) and has very small dust- and gas-disks, of similar radii. This might be due to a companion truncating the disk.

\textit{2MASS J16111742-1918285 (Upper Scorpius 4, $\text{inc}=79.3^{\circ}$, $R_{90\%}$ upper limit of $56$ au, $R_{CO, 90\%}= 47$ au).} This disk is marginally resolved in the visibility plane. The residuals show no significant extra emission. 

\textit{2MASS J16145026-2332397 (Upper Scorpius 5, BV Sco, $\text{inc}=15^{\circ}$, $R_{90\%}$ upper limit of $19$ au, $R_{CO, 90\%}=30$ au).}  This disk is marginally resolved in the visibility plane. The residuals show no significant extra emission. It has a very small gas-disk. \citet{AGEPRO_XI_gas_disk_sizes} proposes it could be a binary because of a residual in the \textsuperscript{12}CO gas-fit located outside of the main gas-disk. This binary might be truncating the gas- and dust-disks.

\textit{2MASS J16163345-2521505 (Upper Scorpius 6, $\text{inc}=62.9^{\circ}$, $R_{90\%}=63$ au, $R_{CO, 90\%}=160$ au).} This is a newly identified \textit{transition disk} with a dust-ring at 35 au. The residuals show no significant extra emission. It shows an inner gas cavity up to 30-40 au (Fig. \ref{Plot: Lupus_gas_dust}), hinting a gas-cleared dust-cavity. A potential dust-cavity was speculated in \citet{2016ApJ...827..142B} and \citet{2018ApJ...859...32P}, which is here confirmed.


\textit{2MASS J16202863-2442087  (Upper Scorpius 7, $\text{inc}=32.4^{\circ}$, $R_{90\%}=47$ au, $R_{CO, 90\%}=160$ au).}  This is a newly identified \textit{transition disk} with a dust-ring at 30 au. The residuals show no significant extra emission. In this paper, we have considered the quiescent continuum data only (see \citealp{AGEPRO_XII_mm_var_USco7}). Due to the complex nature of this source, other \textbf{radial profile} solutions are possible to the one reported here. This disk shows some of the strongest \textsuperscript{12}CO residuals in \citet{AGEPRO_XI_gas_disk_sizes}, which might be indicative of a companion. We refer the reader to \citet{AGEPRO_XII_mm_var_USco7} AGE-PRO paper \textbf{for more details on this source}.

\textit{2MASS J16221532-2511349 (Upper Scorpius 8, $\text{inc}=56.24^{\circ}$, $R_{90\%}=27.19$ au, $R_{CO, 90\%}=145$ au).}  This is a newly identified \textit{transition disk} with a dust-ring at 17 au. Residuals show a localized source of emission eastwards at the $5\sigma$ level. This residual is at the outer border of the dust-ring but well within the gas-disk (\citealp{AGEPRO_IV_UpperSco}). Normally, we would expect to see negative \textsuperscript{12}CO residuals at this location due to the marginally resolved vertical structure of the emitting layer. The fact that, in the work of \citet{AGEPRO_XI_gas_disk_sizes}, we see \textsuperscript{12}CO residuals with lower contrast on this side of the disk compared to the opposite side could be because emission was compensated for by a $5\sigma$ positive peak from a companion. Hence, we propose this source of emission as a possible companion for this source.

\textit{2MASS J16082324-1930009 (Upper Scorpius 9, $\text{inc}=75.10^{\circ}$, $R_{90\%}=47.80$ au, $R_{CO, 90\%}=181$ au).} This is a substructured disk with a shoulder in the intensity radial profile at 20-40 au. The residuals show no significant extra emission. This disk shows an inner gas cavity extending to 10 au, \textbf{which might suggest an unresolved dust cavity at the 10 au scale} (Fig. \ref{Plot: Lupus_gas_dust}). 

\textit{2MASS J16090075-1908526 (Upper Scorpius 10, $\text{inc}=49.14^{\circ}$, $R_{90\%}=42.0$ au, $R_{CO, 90\%}=75$ au).} This is a substructured disk with a shoulder in the intensity radial profile at 20-35 au. This shoulder-like emission in the visibility model resolves into a faint ring when observed at high angular resolution (Fig. \ref{Plot: HR_comp}). Residuals show two asymmetric localized sources of emission northeast and southwest. These could be due to an inaccurate disk inclination or phase center (\citealp{2021ApJ...916...51A}). However, the gas residuals look correct (\citealp{AGEPRO_XI_gas_disk_sizes}), and the shape of these dust residuals are correlated with the shape of the gas residuals. We hence believe these residuals are tracing real substructures.

\bibliography{agepro_Lupus.bib}{}
\bibliographystyle{aasjournal}

\end{document}